\documentclass[journal=mamobx,manuscript=article]{achemso}
\setkeys{acs}{articletitle=true}

\SectionNumbersOn

\usepackage[version=3]{mhchem} 
\usepackage[T1]{fontenc} 

\usepackage{color}
\usepackage{gensymb}

\author{Jack F. Douglas}
\email{jack.douglas@nist.gov}
\affiliation{Materials Science and Engineering Division, National Institute of Standards and Technology, Gaithersburg, Maryland 20899, United States}

\author{Wen-Sheng Xu}
\email{wsxu@ciac.ac.cn}
\affiliation{State Key Laboratory of Polymer Physics and Chemistry, Changchun Institute of Applied Chemistry, Chinese Academy of Sciences, Changchun 130022, P. R. China}

\title{Equation of State and Entropy Theory Approach to Thermodynamic Scaling in Polymeric Glass-Forming Liquids}

\abbreviations{IR,NMR,UV}

\begin{document}

\newpage

\begin{tocentry}

 \centering
 \includegraphics[height=3.25cm]{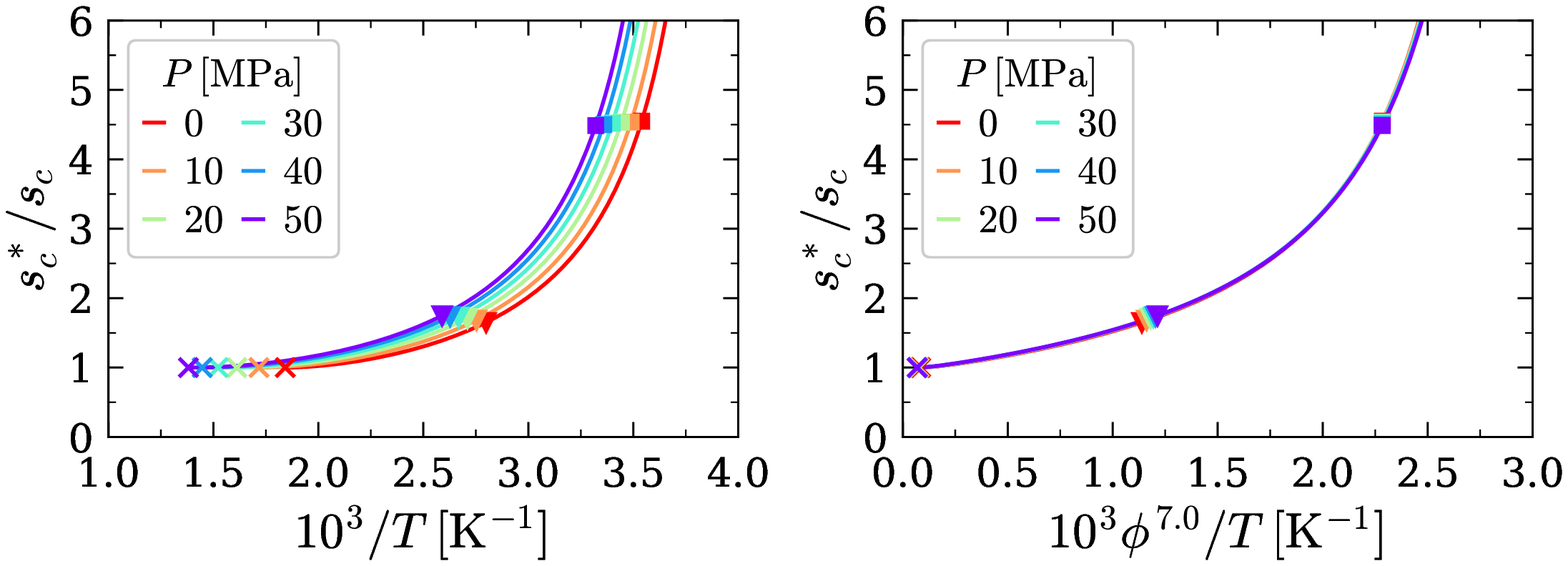}


\end{tocentry}

\newpage

\begin{abstract}
Numerous experimental and computational studies have established that most liquids seem to exhibit a remarkable, yet poorly understood, property termed `thermodynamic scaling' in which the structural relaxation time $\tau_{\alpha}$, and many other dynamic properties, can be expressed in terms of a ‘universal’ reduced variable, $TV^{\gamma}$, where $T$ is the temperature, $V$ is the material volume, and $\gamma$ is a scaling exponent describing how $T$ and $V$ are linked to each other when either quantity is varied. Here, we show that this scaling relation can be derived by combining the Murnaghan equation of state (EOS) with the generalized entropy theory (GET) of glass formation. In our theory, thermodynamic scaling arises in the non-Arrhenius relaxation regime as a scaling property of the fluid configurational entropy density $s_c$, normalized by its value $s_c^*$ at the onset temperature $T_A$ of glass formation, $s_c / s_c^*$, so that a constant value of $TV^{\gamma}$ corresponds to a \textit{reduced isoentropic} fluid condition. Molecular dynamics simulations on a coarse-grained polymer melt are utilized to confirm that the predicted thermodynamic scaling of $\tau_{\alpha}$ by the GET holds both above and below $T_A$ and to test whether the extent $L$ of stringlike collective motion, normalized its value $L_A$ at $T_A$, also obeys thermodynamic scaling, as required for consistency with thermodynamic scaling. While the predicted thermodynamic scaling of both $\tau_{\alpha}$ and $L/ L_A$ is confirmed by simulation, we find that the isothermal compressibility $\kappa_T$ and the long wavelength limit $S(0)$ of the static structure factor do not exhibit thermodynamic scaling, an observation that would appear to eliminate some proposed models of glass formation emphasizing fluid `structure' over configurational entropy. It is found, however, that by defining a low temperature hyperuniform reference state, we may define a compressibility relative to this condition, $\delta \kappa_T$, a transformed dimensionless variable that exhibits thermodynamic scaling and which can be directly related to $s_c / s_c^*$. Further, the Murnaghan EOS allows us to interpret $\gamma$ as a measure of intrinsic anharmonicity of intermolecular interactions that may be directly determined from the pressure derivative of the material bulk modulus. Finally, we show that many phenomenological relationships related to the glass formation and melting of materials can be understood based on the Murnaghan EOS and thermodynamic scaling.
\end{abstract}

\newpage

\section{\label{Sec_Intro}Introduction}

The thermodynamics and dynamics of liquids and solids at equilibrium generally depend on the interplay of inertial dynamics deriving from the thermal energy of the molecules and structural effects associated with emergent density correlations associated with the local tendency of the molecules to form low energy particle configurations. The delicate balance between the kinetic and potential energy contributions to the fluid properties can be altered by the application of an applied pressure $P$ or other type of applied stress. It is the \textit{coupling} of these energetic contributions that somehow leads to the thermodynamic scaling of dynamic properties, such as the structural relaxation time $\tau_{\alpha}$, in terms of a reduced variable involving a product of the temperature $T$ times the volume $V$ to a power $\gamma$ that quantifies the coupling effect, namely, $TV^{\gamma}$.~\cite{2005_RPP_68_1405, 2010_Mac_43_7875, Book_Roland, Book_Paluch} This scaling phenomenon can be more precisely referred to as the `power-law density-temperature scaling', but we adopt the more notationally concise term `thermodynamic scaling', which is also consistent with conventional terminology in experimental studies of this reduced variable description of the dynamics of glass-forming (GF) materials.~\cite{2005_RPP_68_1405, 2010_Mac_43_7875, Book_Roland, Book_Paluch}

To quantitatively describe the thermodynamic scaling of dynamic properties, we must have an equation of state (EOS) that describes how $P$ alters the volume $V$ or density $\rho$ of the material. Such an EOS is expected to provide insight into the material parameter, $\gamma$. An EOS formulation also offers the prospect of developing a universal reduced variable description for properties of the special classes of fluids following Pitzer’s early reasoning and approach to the reduced properties of liquids and gases near their critical points.~\cite{1939_JCP_7_583, 1955_JACS_77_3427, 1955_JACS_77_3433, 2007_JCP_127_224901} In the present work, however, we are interested in the properties of fluids in the low $T$ regime where the fluid compressibility approaches a vanishing small value upon cooling~\cite{2018_PR_745_1, 2003_PRE_68_041113, 2009_JSM_12_P12015, 2016_PRE_94_012902, 2017_PRL_119_136002} rather than diverging upon heating so that a more appropriate reference condition must be chosen for the EOS that addresses the strong interparticle interactions in highly condensed materials at low temperatures and high pressures.

Although linear elasticity provides an adequate framework for describing the dynamics and thermodynamics of materials at low $P$,~\cite{Book_Sokolnikoff, Book_Landau, Book_Slaughter} the relatively incompressible nature of liquids in comparison to gases means that relatively large changes in $P$ are required to impart large changes in the density of condensed materials. Thus, we require a theory that allows us to go beyond the regime where simple linear elasticity is valid. The urgent need for such an EOS theory for modeling geophysical phenomena~\cite{1952_JGR_57_227, 1954_JGR_59_471, 1956_Nature_178_1249, 1957_JATP_10_84} and for understanding high-impact phenomenon~\cite{1938_JCP_6_372} in the 1930s led to the required theoretical framework.

Murnaghan~\cite{1937_AJM_59_235, 1944_PNAS_30_244, 1938_JAP_9_279, 1952_JRMA_1_125} formulated the `integrated linear theory of finite strain', a theoretical framework that naturally resembles the formulation of rubber elasticity from a continuum mechanics perspective.~\cite{1948_PTRSA_241_379} This framework simplifies when combined with assumptions drawn from the Debye theory of solids and further assumptions drawn from the works of Gr{\"u}neisen~\cite{1955_JCP_23_1925} and Gilvarry~\cite{1956_PR_102_331, 1957_JAP_28_1253} quantifying the anharmonic interactions characteristic of real materials. We discuss these anharmonic interactions and their quantification below as these interactions play a central role in determining the thermodynamic scaling exponent, $\gamma$.

Murnaghan’s general EOS for condensed materials subjected to large pressure changes takes the remarkably simple form,~\cite{1944_PNAS_30_244, Book_Murnaghan, 1995_IJT_16_1009}
\begin{equation}
	\label{Eq_Murnaghan}
	P = (1/\gamma_M) B_o [(V_o/V)^{\gamma_M} - 1],
\end{equation}
where the `Murnaghan exponent' $\gamma_M$ was originally interpreted as a phenomenological material constant quantifying the $P$-$V$ relation, reminiscent of the adiabatic expansion law of gases, and $B_o$ is the bulk modulus $B$ at some reference condition indicated by the subscript `$o$', a convention which we also adopt for other quantities in the present paper. For notational economy, eq~\ref{Eq_Murnaghan} is termed the ME. $B_o$ is the reciprocal of the isothermal compressibility $\kappa_{T,o}$ at a chosen reference condition, which is evidently the key thermodynamic property in the above relationship, along with $\gamma_M$. Later works provided a more fundamental interpretation of the ME. In particular, Gilvarry~\cite{1956_PR_102_331, 1957_JAP_28_1253} showed that the EOS of the Debye theory of solids corresponded to a specialization of the ME in which $\gamma_M \equiv \gamma_{\mathrm{Debye}} = 1/3$ and that the ME is a natural extension of the Debye EOS to account for anharmonic intermolecular interactions. Initially, anharmonicity was conceived of in terms of the extent to which the interparticle potential deviated from a quadratic form, but later it became appreciated that the normal modes of crystalline materials can exhibit deviations from the ideal harmonic intermolecular interactions associated with many-body interactions associated with bonding and packing interactions in complex molecules such as polymers, as we discuss below. This alternative type of anharmonicity, which we may term `configurational anharmonicity', can be highly variable in molecular fluids even if the molecules share essentially the same local potential describing their van der Waals interactions, as is often the case in polymeric materials. In particular, configurational anharmonicity is altered by any factor that influences molecular packing of molecules such as molecular rigidity, monomer shape, cohesive interaction strength, and chemical heterogeneity. We also show below that the generalized entropy theory (GET)~\cite{2008_ACP_137_125} enables the analytic calculation of measures of configurational anharmonicity and provides a framework for understanding variations in this type of anharmonicity from a molecular standpoint. We return to a discussion of potential anharmonicity in Section~\ref{Sec_Anharmonicity}, a phenomenon of particular significance for metals and other solids composed of inorganic atomic species where the interatomic species can be quite different from materials with van der Waals interactions, the main concern of the present paper. Although the ME can be derived from specialized models of crystalline solids with non-quadratic local potentials, its derivation from continuum mechanics of materials subjected to large deformations makes this EOS suitable for describing materials for which linear elasticity does not apply. In practice, this means most real solid materials.

In the original derivation of the ME,~\cite{1944_PNAS_30_244, Book_Murnaghan} $B(T)$ was simply expanded in a Taylor series about $P = 0$ and truncated after the leading first-order term and then integrated to obtain eq~\ref{Eq_Murnaghan} where $\gamma$ equals $\partial B / \partial P$ evaluated at $P = 0$, as discussed below. Despite the remarkable simplicity of this continuum theory derivation of the ME, this relation is supported by a large body of measurements.~\cite{1966_JPCS_27_547, 2009_SM_61_453, 2001_APL_79_3947, 2006_ML_60_831} Here, we point out the significance of the $P$ dependence of $B$ in quantifying the strength of `anharmonic interactions' in solid materials,~\cite{1986_PRB_33_2380} as evidenced by deviations from linear elasticity theory in macroscopic elasticity and thermodynamic measurements.

As noted before, real condensed materials normally exhibit significant deviations from the idealized Debye theory arising from the inherently \textit{anharmonic} nature of interparticle interactions in the strongly condensed state.~\cite{1957_AP_1_77} The practical consequences of these interactions are apparent in observations of the thermal and electrical conductivity, as well as in many other thermodynamic and dynamic properties of condensed materials.~\cite{1955_PPSB_68_951, 1955_PPSB_68_957} Gr{\"u}neisen~\cite{1955_JCP_23_1925} introduced a widely utilized measure of anharmonicity based on a consideration of the thermal expansion coefficient $\alpha_P$, the property that perhaps best provides the physical expression of these anharmonic interactions. Note that crystalline materials composed of particles interacting through ideal harmonic interactions exhibit no thermal expansion upon heating, and moreover, their stiffness is independent of $T$.~\cite{1994_PSS_181_81} In particular, Gr{\"u}neisen introduced his famous anharmonicity parameter $\gamma_G$ through the relationship,~\cite{1955_JCP_23_1925}
\begin{equation}
	\label{Eq_GammaG}
	\gamma_G \equiv \alpha_P V / \kappa_T C_V,
\end{equation}
where $C_V$ is the constant volume specific heat. Later, it was realized that the above definition could be formally obtained from the Debye theory by taking the lattice normal mode frequencies to scale with the material volume $V$ to the power $\gamma_G$. While no unique measure of anharmonicity exists even in crystalline materials, Gr{\"u}neisen’s definition for $\gamma_G$ has been established as a practical measure of anharmonicity. Of course, this definition of anharmonicity does not consider \textit{why} the phenomenological scaling of the vibrational frequencies should scale as $V^{\gamma_G}$, as in the continuum theory definition of anharmonicity through the ME. Barker~\cite{1963_JAP_34_107} has emphasized from general dimensional analysis arguments that the dimensionless ratio of properties defining $\gamma_G$ should make this quantity suitable for describing anharmonicity in any condensed material, regardless of whether the material is crystalline or not, and we adopt this perspective in the present work.

Given the manifold sources of `anharmonicity' arising in real condensed materials, the abstract nature of the model assumptions underlying eq~\ref{Eq_GammaG} is probably a strength rather than a weakness in this definition of material anharmonicity. It is also stressed that the derivation of the ME by Gilvarry~\cite{1956_PR_102_331, 1957_JAP_28_1253} assumes that the Poisson ratio $\nu$ of the material is constant, while one may expect both $\nu$ and $\gamma_G$ to vary with $T$~\cite{1966_RMP_38_669, 2003_JPCM_15_603} and generally with thermodynamic conditions in real materials. As we shall discuss below, however, $\gamma_G$ is often found to weakly depend on $T$ in many real materials so that the $T$ variation of $\gamma_G$ can be reasonably neglected, at least in particular ranges of $T$.

Gilvarry~\cite{1956_PR_102_331, 1957_JAP_28_1253} showed that the `Murnaghan exponent' $\gamma_M$ can be directly interpreted in terms of Gr{\"u}neisen’s anharmoncity parameter, $\gamma_G$. Specifically, in a generalized EOS to incorporate anharmonicity of intermolecular interactions, Gilvarry~\cite{1956_PR_102_331, 1957_JAP_28_1253} found $\gamma_M$ to obey the following relation,
\begin{equation}
	\label{Eq_GammaM}
	\gamma_M = 2\gamma_G + 1/3.
\end{equation}
This equation is sometimes called the `Slater equation' in the recent literature, because Slater~\cite{1968_JGR_73_5187, Book_Slater} emphasized its importance. Since the ME reduces to the Debye EOS in the absence of anharmonicity (i.e., $\gamma_G = 0$),~\cite{1957_AP_1_77} the difference, $\Delta \gamma \equiv \gamma_M - 1/3$, provides a measure of anharmonicity of interparticle interactions from the above generalized EOS. Numerous experimental studies have shown that the ME and its extensions to address attractive and repulsive contributions to the intermolecular potential provide a markedly good description of the properties of diverse condensed materials so that we have a framework of sufficient generality to address the EOS relationships in polymeric and other GF fluids in their condensed state. For specificity, we note the excellent work of Anderson~\cite{1966_JPCS_27_547} for metallic and ionic crystalline materials and precise EOS measurements by Wang and coworkers~\cite{2009_SM_61_453, 2001_APL_79_3947, 2006_ML_60_831} for metallic GF materials. These works show that the ME can give an impressively accurate description of the $P$-$V$ relationships in real condensed materials without adjustable parameters. The adaptation of the ME to polymer materials has been discussed in a previous work by Barker,~\cite{1967_JAP_38_4234} and we consider some experimental comparisons to the ME below based on measurements on common polymer materials.

Now, we consider materials in the reference state specified by $V_o$ and $P_o$. If we apply a new pressure to the material, we can obtain the alternative form for condensed polymer materials upon rearrangement of the ME,
\begin{equation}
	\label{Eq_VoV}
	V_o/V = [1 + \gamma_M \kappa_{T,o} (P - P_o)]^{1/\gamma_M}.
\end{equation}
While it is natural to adopt $P = 0$ as a reference condition, as implicit in eq~\ref{Eq_Murnaghan}, we may just as well expand $B$ in a Taylor series about any reference pressure $P_o$, defining a different reference state. Cook and Rogers~\cite{1963_JAP_34_2330} later derived the ME by a different route and their work is of interest for understanding the relation between $\gamma_M \kappa_{T,o}$ and the material cohesive energy density and internal pressure. The work of Cook and Rogers~\cite{1963_JAP_34_2330} also well illustrates that deviations from the ME arise near phase transitions, such as crystal melting and insulating-conducting or magnetic phase transitions, which should come as no surprise since such transitions can lead to sharp changes in the volume or cohesive energy density. `Kinks' in the EOS near thermodynamic transitions have independent interests in relation to the existence of such transitions in GF liquids, which we briefly consider below.

In practice, there is normally little change in the properties of condensed materials under ambient pressure in comparison to those at $P = 0$, and many simulations are performed at $P = 0$ for this reason (e.g., see refs~\citenum{2020_Mac_53_4796, 2020_Mac_53_6828, 2020_Mac_53_9678}). We then arrive at the relation,
\begin{equation}
	\label{Eq_VoVInv}
	(V_o/V)^{\gamma_M} = (\rho/\rho_o)^{\gamma_M} = 1 + \gamma_M \kappa_{T,o} P \equiv 1 + P/P^*,
\end{equation}
where we have defined a `characteristic pressure' $P^*$ for the material. The regime with $P/P^* \gg 1$ defines a high $P$ regime where thermodynamic properties and relaxation times depend strongly on $P$ and the regime with $P/P^* \ll 1$ corresponds to a low $P$ regime where thermodynamic properties and relaxation times are close to those at ambient pressure. The $T$ dependence of the EOS parameter $\gamma_M$ may be inferred from the original continuum dynamic mechanics `derivation' of eq~\ref{Eq_Murnaghan} by Murnaghan,~\cite{1944_PNAS_30_244, Book_Murnaghan} which can be immediately deduced from the assumption that the bulk modulus $B$ varies linearly with $P$,
\begin{equation}
	\label{Eq_B}
	B = B_o + (dB / dP)_o P,\ B \equiv 1/\kappa_T,
\end{equation}
where consistency requires, 
\begin{equation}
	\gamma_M = (dB / dP)_o.
\end{equation}

Experimental estimates of $(dB / dP)_o$ can be conveniently obtained from the well-known Bridgeman equation~\cite{1955_JCP_23_1925, 1966_RMP_38_669, 1966_JPCS_27_547} describing the change of the volume $V$ from its initial reference volume $V_o$ upon the application of a change in the external pressure $P$,
\begin{equation}
	\label{Eq_VVo}
	V / V_o = 1 - (P/B_o) + a_1(P/B_o)^2 + \mathcal{O}[(P/B_o)^3],\ a_1 = [1 + (dB / dP)_o]/2,
\end{equation}
where the higher-order terms $a_i$ are specified by Macdonald~\cite{1966_RMP_38_669} and Anderson.~\cite{1966_JPCS_27_547} However, the higher-order terms in the above Bridgeman equation are rarely reported except in applications involving extremely large $P$. The phenomenological Tait equation is obtained by the following approximation,~\cite{1976_JAP_47_5201}
\begin{equation}
	\label{Eq_Tait}
	- (V - V_o) / V_o P = 1 / (A + DP).
\end{equation}
$A$ and $D$ are often treated as phenomenological constants in applications of this formula, and Cho~\cite{1988_JAP_64_4236} has shown that the ME is recovered for a particular choice of these constants. The above conventional form of the Tait equation is so closely related to the ME that the ME is sometimes described as being the `modified Tait equation',~\cite{1966_RMP_38_669} despite the fact that the ME has a sounder theoretical basis through its derivation by Gilvarry~\cite{1956_PR_102_331, 1957_JAP_28_1253} as an extension of the Debye EOS to incorporate anharmonic interactions.

One of the interesting experimental regularities in measurements on anharmonic interactions in metallic materials, noted first by Slater in the case of metallic materials,~\cite{1940_PR_57_744} is that $a_1$ in eq~\ref{Eq_VVo}, which links $B_o$ to the nonlinear elasticity term, takes a value close to $2.5 \pm 0.5$ for a range of common metals of importance from a manufacturing standpoint, corresponding to $(dB/ dP)_o \approx 4$. In a later careful experimental study, Anderson~\cite{1966_JPCS_27_547} found the values of $(dB/ dP)_o$ for a range of metals, salts, and minerals to be roughly consistent with the earlier work of Slater,~\cite{1940_PR_57_744} $(dB/ dP)_o \approx 3.5$ to $7.0$. Assuming the ME, a `pressure coefficient' of $(dB/ dP)_o \approx 4$ then corresponds to $\gamma_M \approx 4$. Barker~\cite{1967_JAP_38_4234} extended this pioneering analysis of the EOS of solid materials to encompass newly developed polymer materials where the first and second pressure coefficients $a_1$ and $a_2$ for polymeric materials were estimated to take `typical' values in the corresponding ranges, $a_1 = 4.1 \pm 0.1$ and $a_2 = 8.8 \pm 0.2$. This corresponds to a rough estimate of $\gamma_M = 7.0 \pm 0.2$ for polymer materials, a value that we will find below to be remarkably consistent with estimates of the exponent $\gamma$ for thermodynamic scaling governing the relaxation dynamics of our GF liquid from both the analytic GET~\cite{2008_ACP_137_125} and molecular dynamics (MD) simulations. McGowan~\cite{1970_Polymer_11_436} has also considered the EOS of polymer materials based on the ME, where emphasis was given to the mass dependence of the EOS parameters of the ME. Further studies along this line would be quite interesting.

Macdonald~\cite{1966_RMP_38_669} has reviewed estimates of $\gamma_M$ for molecular fluids including water, hydrocarbon fluids, and metals, along with the $T$ dependence of $\gamma_M$. The precise agreement between the measurements of Anderson~\cite{1966_JPCS_27_547} and the ME for such a wide range of materials is highly encouraging regarding the general applicability of the ME, at least as a good first approximation, to diverse condensed materials. As a general trend, we expect that anharmonic effects, as quantified by $\gamma_M$, tend to become larger with increasing molecular structural and energetic complexity.

At a purely mathematical level, the derivation of the ME simply requires the integration~\cite{1966_RMP_38_669, 1966_JPCS_27_547, 1975_JCP_63_3379} of a linear equation for $B(P)$. It is evident that eq~\ref{Eq_B} neglects the higher-order terms in the $P$ expansion of $B$ whose incorporation has been the focus in refinements of the ME in applications to astrophysical and geophysical phenomena where $P$ can be extremely large,~\cite{2019_Minerals_9_562} but these corrections are often relatively small, at least in small-molecule liquids,~\cite{1975_JCP_63_3379} and the $T$ dependence of $(dB/ dP)_o$ of simple molecules is often modest.~\cite{1966_RMP_38_669, 1993_CIT_65_81} The reader is warned, however, that measurements of the anharmonicity parameter $\gamma_G$, based on the defining equation in eq~\ref{Eq_GammaG}, can be appreciably dependent on $T$ in polymers~\cite{1967_JAP_38_4234} so that the $T$ dependence of the $P$ derivative $(dB / dP)_o$ in polymers of different types should be carefully examined. It is also notable that Poisson ratio $\nu$,~\cite{2011_NatMat_10_823} which is assumed to be constant in the Gilvarry’s more fundamental formulation of the ME,~\cite{1956_PR_102_331, 1957_JAP_28_1253} varies in polymeric and other GF materials. We might naturally expect $\gamma_G$, and thus $\gamma_M$, to be $T$ dependent. We return to this discussion in Section~\ref{Sec_Result}.

\begin{figure*}[htb!]
	\centering
	\includegraphics[angle=0,width=0.475\textwidth]{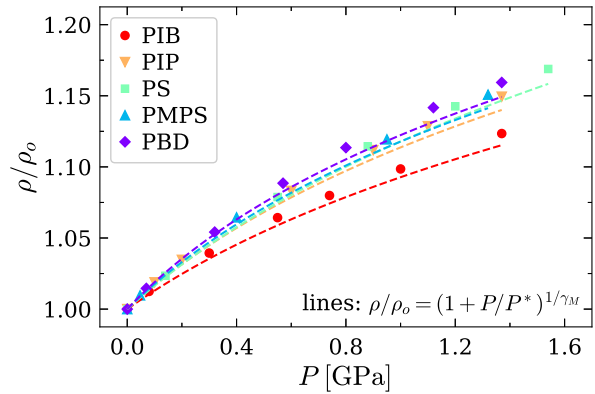}
	\caption{\label{Fig_Rho}Pressure dependence of the density for polymer materials. The plot shows the reduced density $\rho / \rho_o$ versus pressure $P$ for different polymers at a temperature of $T = 140$ K. $\rho_o$ is the density at ambient pressure. PIB, PIP, PS, PMPS, and PBD are short for polyisobutylene, polyisoprene, polystyrene, poly(methylphenyl siloxane), and 1,2-polybutadiene, respectively. Lines correspond to the equation: $\rho / \rho_o = (1 + P/P^*)^{1/\gamma_M}$. Experimental data were taken from ref~\citenum{2008_PRB_78_134201}, where the fitted parameters and experimental uncertainties are discussed.}
\end{figure*}

As many of the references just mentioned are not recent, and oriented to metallic and inorganic materials of geophysical interest, it is worth pointing out the applicability of the ME to common polymeric and other GF materials of relatively recent interest, such as metallic glasses, as a basic `reality check' given the importance of the ME in our development below. In Figure~\ref{Fig_Rho}, we show the variation of the reduced density $\rho / \rho_o$ with $P$ for different polymers measured by the Sokolov group,~\cite{2008_PRB_78_134201} including polyisobutylene (PIB), polyisoprene (PIP), polystyrene (PS), poly(methylphenyl siloxane) (PMPS), and 1,2-polybutadiene (PBD). The ME exponents $\gamma_M$ were found to equal $9.16$, $8.66$, $7.48$, $8.73$, and $9.12$ for PIB, PIP, PS, PMPS, and PBD, respectively, and the isothermal compressibility parameter $P^*$ was tabulated for these polymers by Sokolov and coworkers.~\cite{2008_PRB_78_134201} These estimates are to be compared with Baker's earlier estimates of a `typical' value of $7.0 \pm 0.2$ for polymer materials,~\cite{1967_JAP_38_4234} as mentioned above. We shall see below that these values of $\gamma_M$ are also comparable to those in our GET calculations and MD simulations of model polymeric GF liquids. We point out that the ME also provides an excellent description of relatively recent measurements on metallic GF materials.~\cite{2009_SM_61_453, 2001_APL_79_3947, 2006_ML_60_831}

As a final point in this section, we mention that the EOS discussed above has implications for understanding many other important properties of condensed materials under high pressure conditions, such as the refractive index~\cite{1980_JCP_72_4063} and dielectric properties~\cite{1962_JCP_36_3062} and the pressure dependence of the melting temperature~\cite{1956_PR_102_325, 1967_PR_161_613, 1976_JAP_47_5201, 1966_PR_151_668} and the glass transition temperature,~\cite{2012_PRE_86_041502, 2017_PRL_119_025702} an application that we briefly consider in the present paper.

\section{Methods}

\subsection{\label{Sec_GET}Generalized Entropy Theory}

The GET essentials have been described in our recent works,~\cite{2020_Mac_53_7239, 2020_Mac_53_9678, 2021_Mac_XX_XXXX} so we only present a brief summary here. The GET enables analytically investigating the influence of basic molecular parameters on polymer glass formation.~\cite{2008_ACP_137_125} The basic idea is to merge the lattice cluster theory (LCT)~\cite{1998_ACP_103_335, 2014_JCP_141_044909} for the thermodynamics of polymer systems and the Adam-Gibbs (AG) relation~\cite{1965_JCP_43_139} between the segmental structural relaxation time and the configurational entropy. The LCT yields an analytic expression for the Helmholtz free energy $f$ as a function of temperature $T$, polymer filling fraction $\phi$, molecular mass $M$, microscopic cohesive energy parameter $\epsilon_{lm}$, bending energy parameter $E_b$, and a set of geometrical indices that reflect the size, shape, and bonding patterns of the monomers.

Computations of the characteristic properties of polymer glass formation are achieved by first analyzing the $T$ dependence of the configurational entropy per lattice site, $s_c = - \left. \partial f / \partial T \right|_{\phi}$. For instance, the onset temperature $T_A$ signals the onset of non-Arrhenius behavior of $\tau_{\alpha}$ and can be determined by a temperature corresponding to the maximum $s_c^*$ of $s_c$. The crossover temperature $T_c$ separates two regimes of $T$ with qualitatively different dependences of $\tau_{\alpha}$ on $T$ and is estimated from $\partial^2 (Ts_c)/\partial T^2 = 0$.~\cite{2008_ACP_137_125} The determination of the glass transition temperature $T_{\text{g}}$ follows its operational definition based on the condition, $\tau_{\alpha}(T_{\text{g}}) = 100$ s, where $\tau_{\alpha}$ is computed from the AG relation,~\cite{1965_JCP_43_139}
\begin{equation}
	\label{Eq_AG}
	\tau_{\alpha} = \tau_o \exp\left( \frac{\Delta G_o}{k_BT} \frac{s_c^*}{s_c} \right).
\end{equation}
The GET assumes that the high $T$ vibrational prefactor equals $\tau_o = 10^{-13}$ s, a typical experimental estimate for polymers.~\cite{2003_PRE_67_031507} $\Delta G_o$ is the activation free energy at high $T$, which is anticipated from transition state theory (TST)~\cite{1941_CR_28_301, Book_Eyring} to contain both enthalpic $\Delta H_o$ and entropic $\Delta S_o$ contributions, i.e., $\Delta G_o = \Delta H_o - T \Delta S_o$. Motivated by the approximation made by AG~\cite{1965_JCP_43_139} that the entropic contribution to the activation free energy is negligible and the experimental evidence~\cite{2008_ACP_137_125} that $\Delta H_o \approx 6 k_B T_c$, the GET simply assumes that $\Delta G_o = 6 k_B T_c$. The reasoning for such a relationship between $\Delta H_o$ and $T_c$ has been discussed at length in ref~\citenum{2008_ACP_137_125}, and we refer the reader to this reference for details. We note that $\Delta H_o$ can alternatively be estimated from experiment or simulation,~\cite{2015_JCP_143_144905} but the approximation adopted by the GET has the advantage that $\tau_{\alpha}$ for polymer melts can be predicted based solely on molecular and thermodynamic parameters required to describe their thermodynamics. Finally, the characteristic temperature $T_0$ demarking the `end' of glass formation,~\cite{2008_ACP_137_125} along the fragility index $D$, can be obtained from the Vogel-Fulcher-Tammann (VFT) equation,~\cite{1921_PZ_22_645, 1925_JACS_8_339, 1926_ZAAC_156_245}
\begin{equation}
	\label{Eq_VFT}
	\tau_{\alpha} = \tau_{0}\exp\left(\frac{DT_0}{T-T_0}\right),
\end{equation}
where $\tau_{0}$ is a prefactor. The VFT equation has been found to describe the $T$ variation of $\tau_{\alpha}$ calculated from the GET~\cite{2008_ACP_137_125} rather well over a $T$ range above $T_{\text{g}}$, but below $T_c$. Notably, the GET allows for the direct calculation of $\tau_{0}$, $D$, and $T_0$ in terms of molecular parameters governing the thermodynamics of the fluid.~\cite{2008_ACP_137_125} There is a vibrational contribution to the prefactor $\tau_{0}$ that scales inversely to an average phonon frequency, which is assumed to take a `typical' value in molecular liquids, $10^{-13}$ s, although this time scale can be estimated precisely from the initial decay of the velocity autocorrelation function of the polymer segments or from an instantaneous normal mode analysis.~\cite{2019_JCP_151_184904}

The present paper considers a melt of chains with the structure of polypropylene (PP) under a range of fixed $P$ conditions, where the $P$ range is selected to ensure that thermodynamic scaling holds.~\cite{2013_JCP_138_234501} In all our calculations, the cell volume parameter is $V_{\text{cell}} = 2.5^3 \text{\AA}^3$, the chain length is $N_c = 8000$, the cohesive energy parameter is $\epsilon_{lm}/k_B = 200$ K, and the bending energy parameter is $E_b/k_B = 640$ K. The selected chain length is typical of high molecular masses, for which $\tau_{\alpha}$ remains nearly unchanged with varying $N_c$. The choices of $\epsilon_{lm}$ and $E_b$ yield a thermodynamic scaling exponent that is consistent with the value estimated from our simulations, as shown in Section~\ref{Sec_Result}.

\subsection{\label{Sec_MD}Molecular Dynamics Simulation}

Our simulation study of polymer glass formation is based on a coarse-grained bead-spring model of polymer melts.~\cite{1990_JCP_92_5057, 1986_PRA_33_3628} Details for this model have been given elsewhere,~\cite{2020_Mac_53_6828} along with the simulation methodology, so our description is brief. This model assumes that each polymer chain is composed of a number of connected beads, where bond connectivity along neighboring beads is maintained by the finitely extensible nonlinear elastic (FENE) potential,~\cite{1990_JCP_92_5057, 1986_PRA_33_3628}
\begin{eqnarray}
	\label{Eq_FENE}
	U_{\text{FENE}}(r) = -\frac{1}{2} k_b R_0^2 \ln\left[1 - \left(\frac{r}{R_0}\right)^2\right] + 4 \varepsilon \left[ \left(\frac{\displaystyle \sigma}{\displaystyle r}\right)^{12} - \left(\frac{\displaystyle \sigma}{\displaystyle r}\right)^6 \right] + \varepsilon,
\end{eqnarray}
where $r$ denotes the distance between two beads and $\varepsilon$ and $\sigma$ are the length and energy scales associated with the Lennard-Jones (LJ) potential. The firs term of eq~\ref{Eq_FENE} extends to $R_0$, and the second term has a cutoff at $2^{1/6} \sigma$. Common choices are adopted for the parameters $k_b$ and $R_0$, namely, $k_b = 30 \varepsilon/\sigma^2$ and $R_0 = 1.5 \sigma$. Chain stiffness is controlled by an angle potential with the following form,~\cite{2019_JCP_150_091101}
\begin{eqnarray}
	\label{Eq_Bend}
	U_{\text{bend}}(\theta) = -a_{\theta} \sin^2(b_{\theta} \theta),\ 0 < \theta < \pi/b_{\theta},
\end{eqnarray}
where the bond angle is given by $\theta = \cos^{-1}[(\mathbf{b}_{j} \cdot \mathbf{b}_{j+1})/ (|\mathbf{b}_{j}| |\mathbf{b}_{j+1}|)]$ in terms of the bond vector $\mathbf{b}_{j} = \mathbf{r}_{j} - \mathbf{r}_{j-1}$ between two neighboring beads $j$ and $j-1$. We adopt $b_{\theta} = 1.5$ based on our previous work.~\cite{2020_Mac_53_2754} Nonbonded pair interactions are described by a truncated-and-shifted LJ potential,
\begin{eqnarray}
	\label{Eq_LJ}
	U_{\text{LJ}}(r) = 4 \varepsilon \left[ \left(\frac{\displaystyle \sigma}{\displaystyle r}\right)^{12} - \left(\frac{\displaystyle \sigma}{\displaystyle r}\right)^6 \right] + C(r_{\text{cut}}),\ r < r_{\text{cut}},
\end{eqnarray}
where $C(r_{\text{cut}})$ is a constant to ensure that $U_{\text{LJ}}$ varies smoothly to zero at the cutoff distance $r_{\text{cut}}$. In the present work, we use $r_{\text{cut}} = 2.5\sigma$ to include attractive nonbonded interactions.

Each chain has $120$ beads and the total bead number is $N=12000$ in our polymer system. In accord with our calculations based on the GET, the chain length with $120$ beads lies in a mass regime where the segmental dynamics is insensitive to changes in molecular mass. We describe all quantities and results from MD simulations in standard reduced LJ units. Specifically, length, time, temperature, and pressure are, respectively, given in units of $\sigma$, $\tau$, $\varepsilon/k_B$, and $\varepsilon/\sigma^3$, where $\tau = \sqrt{m\sigma^2/\varepsilon}$ with $m$ being the bead mass and $k_B$ is Boltzmann's constant. We utilize the HOOMD-blue simulation package~\cite{HOOMD_2008, HOOMD_2015, HOOMD_webpage} to perform all our simulations, and details can be found in refs~\citenum{2020_Mac_53_6828} and~\citenum{2020_Mac_53_2754}.

\section{\label{Sec_Result}Results and Discussion}

\subsection{Thermodynamic Scaling in the Arrhenius Relaxation Regime}

We first consider a relationship between dynamics and thermodynamics in the high $T$ Arrhenius regime where TST in its classical form~\cite{1941_CR_28_301, Book_Eyring} is applicable. The application of $P$ modifies the enthalpy $\Delta H_A(P)$ of activation as,
\begin{equation}
	\label{Eq_DHA}
	\Delta H_A(P) = \Delta H_{A}(0) + P V_{\mathrm{act}},
\end{equation}
where $V_{\mathrm{act}}$ is the activation volume and the `A' subscript on $\Delta H_A(P)$ and its zero pressure analog $\Delta H_{A}(0)$ serve as a reminder that our discussion is limited to the Arrhenius regime where $\Delta H_A(P)$ is independent of $T$. As noted by Keyes,~\cite{1958_JCP_29_467} the results of Slater,~\cite{Book_Slater} based on the elasticity theory of Eshelby,~\cite{Book_Eshelby} and the Gr{\"u}neisen's extension of the Debye model of anharmonic solids indicate that $V_{\mathrm{act}}$ can be directly estimated from $\Delta H_{A}(0)$ in terms of $\gamma_M$ and $\kappa_{T,o}$,
\begin{equation}
	\label{Eq_Vact}
	V_{\mathrm{act}} = \gamma_M \kappa_{T,o} \Delta H_{A}(0).
\end{equation}
$\Delta H_A(P)$, in turn, takes the following form,
\begin{equation}
	\Delta H_A(P) = \Delta H_{A}(0) (1 + \gamma_M \kappa_{T,o} P) = \Delta H_{A}(0) (1 + P/P^*) = \Delta H_{A}(0) (V/V_o)^{-\gamma_M},
\end{equation} 
where we have used eq~\ref{Eq_VoVInv}. See ref~\citenum{1962_JCP_37_2785} for a discussion of eq~\ref{Eq_Vact} and its applicability to polymer materials. The effect of anharmonicity on the activation energy for relaxation thus leads to a rescaling of $\Delta H_{A}(P)$ in the $P = 0$ reference state by a $V$-dependent factor. We then see that the dimensionless activation energy $\Delta H_A(P) / k_BT$ scales as,
\begin{equation}
	\label{Eq_DHA2}
	\Delta H_A(P) / k_BT \sim (\epsilon / k_BT) (V/V_o)^{-\gamma_M} \sim (\epsilon / k_BT) (\rho/\rho_o)^{\gamma_M},
\end{equation}
where $\epsilon$ is a cohesive interaction parameter describing the strength of attractive interparticle interactions (e.g., the well depth of intermolecular potential) required to make the $T$ factor dimensionless,~\cite{2016_ACP_161_443, 2016_Mac_49_8355, 2020_Mac_53_9678} which is denoted $\epsilon_{lm}$ in our lattice model. The appropriate dimensionless structural relaxation time and diffusion coefficient in the Arrhenius regime are then a function of $(\epsilon / k_BT) (\rho/\rho_o)^{\gamma_M}$. The existence of this reduced scaling indicates that it is problematic to naively replace $T$ by reciprocal volume fraction~\cite{2009_Nature_462_83} or pressure~\cite{2002_Mac_35_7338} in models or correlative expressions of the $T$ dependence of the relaxation time or diffusion coefficient of molecular GF liquids. This scaling has been confirmed in simulations of model simplified metallic GF liquids in the Arrhenius regime by Tarjus and coworkers,~\cite{2004_JCP_120_6135} where $\gamma_M$ was estimated to be about $4$, as in classical estimates for metals. This is a typical value of $\gamma_M$ in metallic materials and Keyes~\cite{1958_JCP_29_467} took this a general estimate of the proportionality factor in eq~\ref{Eq_Vact} (See Figure 1 in ref~\citenum{1958_JCP_29_467}). Alba-Simionesco et al.~\cite{2004_EPL_68_58} and Dreyfus et al.~\cite{2004_EPJB_42_309} observed that the scaling relation holds over a wide range of $T$ for a wide range of GF liquids. In the next section, we attempt to understand the origin of this unexpected phenomenon.

We remark that the argument for the `activation energy' $\Delta H_A$ by Keyes~\cite{1958_JCP_29_467} is an early variant of the `shoving model'~\cite{2006_RMP_78_953, 2012_JCP_136_224108, 2015_JNCS_407_14} in which $\Delta H_A$ is modeled as being proportional to the shear modulus $G$ based on the continuum theory arguments of Eshelby~\cite{Book_Eshelby} for the energetic cost of displacing molecules in a condensed material as a description of the activation energy. Other authors have suggested that the activation energy should be more appropriately proportional to the bulk modulus $B$.~\cite{1980_PRB_22_3130, 1981_PRB_24_904, 2009_PRE_79_032501} Given that the Poisson ratio has been assumed to be constant in the Gilvarry formulation~\cite{1956_PR_102_331, 1957_JAP_28_1253} of the ME extending the Debye theory and the $P$ dependence on the differential change of the shear modulus, this alternative model of $\Delta H_A$ leads to exactly the same result as noted by Keyes~\cite{1958_JCP_29_467}. Of course, the variation of the Poisson ratio in cooled liquids at low $T$, which has been observed experimentally in PS and other GF liquids, means that we must be prepared to consider a $T$-dependent anharmonicity in the description of cooled liquids in the $T$ regime below the onset temperature $T_A$ for non-Arrhenius relaxation and diffusion. We consider the implications of a $T$-dependent anharmonicity in Section~\ref{Sec_Anharmonicity}.

Parenthetically, we also note that Keyes~\cite{1958_JCP_29_467} suggested a tentative expression for the activation entropy $\Delta S_{A}$ in which the anharmonicity parameter $\gamma_M$ likewise arises, $\Delta S_{A} = \gamma_M \alpha_P \Delta H_{A}$, a relation that draws upon the earlier arguments of Lawson~\cite{1957_JPCS_3_250} and the highly influential work of Wert and Zener~\cite{1949_PR_76_1169}. This relation is quite interesting since any change of molecular structure or interaction that alters the anharmonicity should alter the entropy and enthalpy of activation in a proportionate fashion, a phenomenon termed the `entropy-enthalpy compensation' (EEC).~\cite{2001_CR_101_673, 2006_PPP_69_1145, 2011_EPJB_82_271} Moreover, for materials in the low $T$ solid state, $\alpha_P$ can be taken to be essentially constant, so that we can use the near constancy $\alpha_P T_m$ for crystalline materials and $\alpha_P T_{\mathrm{g}}$ for glassy materials to arrive at an approximate proportionality between $\Delta S_A$ and $1/T_m$ and $1/T_{\mathrm{g}}$, respectively. Such a phenomenology, whose origin is discussed below in Section~\ref{Sec_Implication}, is widely observed in diverse crystalline and glass materials.~\cite{1950_JAP_21_1189, 1951_JAP_22_372, 1959_JCP_31_135, 1980_AM_28_1085, 1997_Polymer_38_1081, 1984_Polymer_25_299, 1993_JAP_74_1597, 1988_JPD_21_1519, 2008_JPCB_112_15980, 2017_JCP_147_154902, 2015_PNAS_112_2966, 2018_NanoLett_18_7441} In particular, it seems suitable to choose $\alpha_P T_{\mathrm{g}} = 0.16$ based on the empirical rule of Boyer and Bondi for polymers, which indicates that $\alpha_P T_{\mathrm{g}}$ lies in the range $0.016$ to $0.19$,~\cite{Book_VanKrevelen} which implies $\Delta S_{A} \approx \Delta H_{A} / T_{\mathrm{g}}$ if we take $\gamma_M$ to have a representative value of about $6$, a value roughly consistent with our results from the GET and MD simulations below for coarse-grained models of polymer fluids. The product $\alpha_P T_{\mathrm{g}}$ is also nearly constant in metallic GF materials~\cite{2008_SM_58_1106} and $\alpha_P T_m$ is well-known to be fairly constant in metallic crystalline materials.~\cite{2010_APL_97_171911} Curiously, the Keyes relation for $\Delta S_{A}$ implies that the compensation temperature correlates with $T_m$ for crystalline materials and $T_{\mathrm{g}}$ for materials having a strong propensity to form glasses. We plan to investigate $\Delta S_{A}$ systematically in the future in view of evidence that this quantity is closely related to the dynamics of condensed materials. We note that the AG model~\cite{1965_JCP_43_139} and the GET~\cite{2008_ACP_137_125} simply neglect the activation entropy, which is arguably a major shortcoming of these models.

\subsection{Thermodynamic Scaling in the Non-Arrhenius Relaxation Regime}

It is not immediately clear whether or not the scaling in eq~\ref{Eq_DHA2} applies in the $T$ regime below the onset temperature $T_A$ for non-Arrhenius relaxation, a regime that is evidently the most practically interesting for polymeric and other GF liquids. We can anticipate that anharmonic interaction effects become more prevalent in the $T$ regime below $T_A$ and that these interactions presumably give rise to collective particle motion underlying the non-Arrhenius relaxation.~\cite{1998_PRL_80_2338, 2013_JCP_138_12A541, 2014_JCP_140_204509, 2006_PRL_97_045502, 2014_NatCommun_5_4163, 2015_JCP_142_234907, 2019_JPCB_123_5935, 2015_PNAS_112_2966, 2016_Mac_49_8355, 2016_MacroLett_5_1375, 2017_Mac_50_2585, 2017_SoftMatter_13_1190, 2019_JCP_150_101103, 2015_JCP_142_164506, 2013_ACP_152_519, 2018_JCP_149_161101, 2013_SoftMatter_9_241, 2011_PRL_106_115702, 2020_JCP_152_054904, 2020_Mac_53_4796, 2020_Mac_53_6828} In Section~\ref{Sec_Anharmonicity}, we show that the growth of collective motion identified in our simulated polymer melts follows an increase in the anharmonic interaction strength, as quantified by the derivative of the bulk modulus with respect to $P$.

An impressive body of evidence from both experiment and simulation indicates that the thermodynamic scaling exponent $\gamma$ obtained from estimates of viscosity and relaxation time lies in the broad range of $0.18$ to $8.5$.~\cite{2005_RPP_68_1405} Notably, thermodynamic scaling appears to hold over the entire $T$ range of glass formation in diverse materials,~\cite{2005_RPP_68_1405, Book_Roland, Book_Paluch} leading to different theoretical rationalizations of this fact. Some authors~\cite{2008_PRL_100_015701, 2009_JCP_131_234504} have suggested that this scaling property derives from a `hidden symmetry' associated with the power-law nature of the pairwise repulsive intermolecular potential that is typical of molecular fluids, along with the assumption that attractive intermolecular interactions can be neglected in fluids in comparison to the harsh repulsive interactions in liquids.~\cite{2008_PRL_100_015701, 2009_JCP_131_234504} Alternatively, Casalini, Roland, and coworkers~\cite{2006_JCP_125_014505, 2007_JPCM_19_205118, 2007_PM_87_459} have emphasized that thermodynamic scaling arises from a scaling property of the fluid configurational entropy, a thermodynamic explanation with no obvious relation to the functional form of the intermolecular potential.~\cite{2008_PRL_100_015701, 2009_JCP_131_234504} Recent simulation studies of numerous GF liquids, including polymeric,~\cite{2014_JCP_141_054904, 2015_JCP_143_194503, 2016_JCP_145_234904} ionic and molecular,~\cite{2011_JCP_134_144507, 2011_JCP_135_164510} and metallic GF liquids,~\cite{2016_JCP_145_104503} have indicated that the pressure-energy correlations associated with the power-law potential and the associated virial theorem linking the potential and kinetic energies of the fluid do not generally exist so that the power-law potential argument cannot be a general explanation for thermodynamic scaling. We also note an instructive study of thermodynamic scaling in polymeric fluids based on the GET~\cite{2013_JCP_138_234501} where the thermodynamic scaling of the segmental structural relaxation time was observed and the scaling exponent $\gamma$ was found to depend on the chain length, chain rigidity, cohesive interaction strength, and side-group length, factors that evidently influence the fragility of glass formation. The GET is based on a lattice model to enable analytically tractable calculations of the configurational entropy density $s_c$, where the intermolecular potential roughly has the form of a square-well potential. The potential is invariant in polymer melts having different molecular parameters that obviously greatly influence the thermodynamic scaling exponent, $\gamma$. This provides unequivocal evidence for the existence and practical importance of configurational anharmonicity on the properties of polymers and other molecular materials. Calculations based on the GET~\cite{2013_JCP_138_234501} also confirmed the approximate inverse relation between $\gamma$ and the constant volume fragility parameter $m_V$ and the nearly linear relation between $m_V$ and its constant pressure fragility counterpart $m$ found experimentally by Casalini and Roland.~\cite{2007_JNCS_353_3936, 2005_PRE_72_031503} These results also strongly support the view of thermodynamic scaling proposed by Casalini and Roland,~\cite{2007_JPCM_19_205118} which emphasizes that thermodynamic scaling reflects the influence of anharmonic interactions on the $T$ dependence of the configurational entropy. We next probe the extent to which thermodynamic scaling applies to the configurational entropy based on the GET and the extent of cooperative exchange motion based on MD simulations to examine the interpretation of Casalini and Roland~\cite{2007_JNCS_353_3936, 2005_PRE_72_031503} more quantitatively. 

Casalini and Roland~\cite{2007_JPCM_19_205118} have previously investigated the thermodynamic scaling of excess entropy estimates of model GF liquids, a quantity that is often taken as an experimental counterpart of the configurational entropy, which cannot be directly measured experimentally, and they found that the thermodynamic scaling exponent for the configurational entropy was somewhat different than that for dynamic properties. It was not clear how much the same conclusion was affected by their approximate estimate of the configurational entropy, however. We can completely avoid the uncertainties in estimates of the configurational entropy in our calculations based on the GET. The MD simulations should be helpful in establishing the extent to which the scaling property of the configurational entropy is transferable to the dynamics of GF liquids.

\subsection{Thermodynamic Scaling based on the GET}

The GET involves a combination of the LCT of polymer thermodynamics, developed initially by Freed and coworkers,~\cite{1998_ACP_103_335, 2014_JCP_141_044909} to address the thermodynamics of multicomponent polymers having nontrivial monomer structure, with the AG relation,~\cite{1965_JCP_43_139} which provides a formal link between the thermodynamics and dynamics of GF materials. We have calculated $s_c$ for many systems in the past, but we never tested whether or not $s_c$ exhibits thermodynamic scaling. The observation of thermodynamic scaling over a wide $T$ range and the consistency of the same scaling with the GET evidently require that thermodynamic scaling should exist. Moreover, the scaling found in the high $T$ Arrhenius regime must match the thermodynamic scaling in the $T$ regime below the onset temperature $T_A$ for non-Arrhenius relaxation and diffusion. This is not at all obviously true, and examining thermodynamic scaling provides a strong test for the GET framework and an opportunity to learn more about this mysterious phenomenon.

\begin{figure*}[htb!]
	\centering
	\includegraphics[angle=0,width=0.975\textwidth]{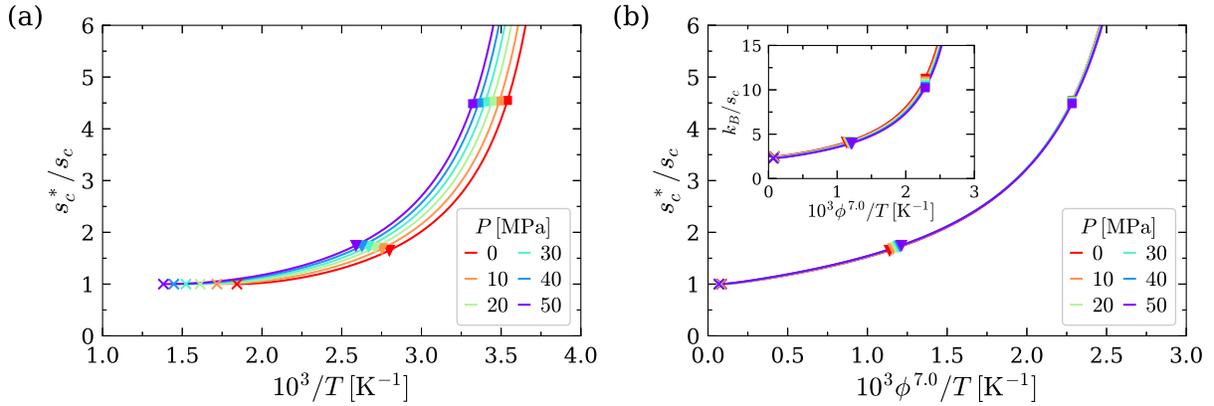}
	\caption{\label{Fig_Sc}Generalized entropy theory test of thermodynamic scaling of the configurational entropy density. (a) Configurational entropy density ratio $s_c^*/s_c$ versus inverse temperature $10^3/T$ for a range of $P$. The cross, triangle, and square symbols indicate the positions of the onset $T_A$, crossover $T_c$, and glass transition temperatures $T_{\mathrm{g}}$ of glass formation, respectively. (b) $s_c^*/s_c$ versus $10^3 \phi^{\gamma}/T$ with $\gamma =7.0$ for this particular polymer model. The exponent $\gamma$ can be tuned by changing molecular parameters, such as chain length, cohesive interaction strength, chain stiffness, monomer structure, etc. The inset in panel (b) shows that $s_c$ by itself \textit{does not} exhibit thermodynamic scaling.}
\end{figure*}

\begin{figure*}[htb!]
	\centering
	\includegraphics[angle=0,width=0.975\textwidth]{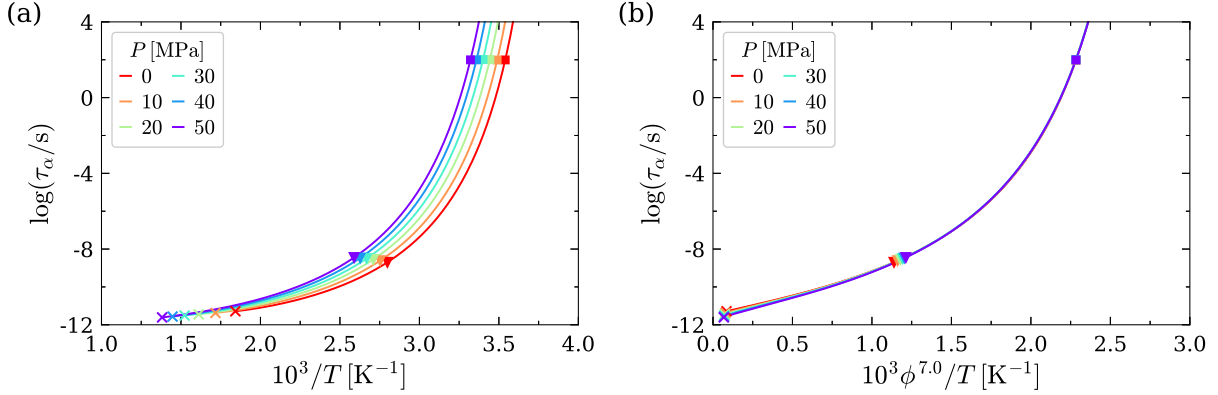}
	\caption{\label{Fig_Tau_GET}Generalized entropy theory test of thermodynamic scaling of the structural relaxation time. (a) Logarithm of the structural relaxation time $\tau_{\alpha}$ versus inverse temperature $10^3/T$ for a range of $P$. The cross, triangle, and square symbols indicate the positions of the onset $T_A$, crossover $T_c$, and glass transition temperatures $T_{\mathrm{g}}$ of glass formation, respectively. (b) $\log \tau_{\alpha}$ versus $10^3 \phi^{\gamma}/T$ with $\gamma =7.0$. Note that the values of $\tau_{\alpha}$ at $T_A$ and $T_c$ are relatively insensitive to $P$ for the range of $P$ investigated.}
\end{figure*}

In Figure~\ref{Fig_Sc}a, we show the configurational entropy density ratio $s_c^*/s_c$ versus $10^3/T$ calculated from the GET for a model of polymer fluids over a wide range of $T$ for a range of fixed $P$ values ranging from $0$ MPa to $50$ MPa. We see that altering $P$ greatly changes the $T$ dependence of $s_c^*/s_c$.

We now attempt to rescale all our data for $s_c^*/s_c$ by invoking the existence of thermodynamic scaling in Figure~\ref{Fig_Sc}b. All our curves for $s_c^*/s_c$ indeed reduce to a remarkably universal master curve in terms of $\phi^{\gamma} / T$ with $\gamma = 7.0$, where this exponent is dependent on the particular polymer model.~\cite{2013_JCP_138_234501} Note that $\phi$ is the \textit{dimensionless} filling fraction defined by the ratio of the total number of united-atom groups to the total number of lattice sites. In the GET, the density, analogous to $\rho$ in our simulations, may be defined by $\phi/V_{\text{cell}}$ with $V_{\text{cell}}$ being the volume of a single lattice site, but we adopt $\phi$ in our analysis for convenience. Evidently, the GET gives rise to a thermodynamic scaling of $s_c^*/s_c$. Thermodynamic scaling for $s_c^*/s_c$ also holds if we vary molecular parameters, such as chain length, cohesive interaction strength, chain stiffness, and monomer structure, and the exponent $\gamma$ clearly depends on the polymer molecular structure and other factors that influence molecular packing in space and the `anharmonicity' of the intermolecular interactions.~\cite{2013_JCP_138_234501} We emphasize again that these anharmonic interactions are of a many-body nature rather than being related to the form of the local pair potential. However, thermodynamic scaling does not hold for $s_c$ itself, as shown in the inset of Figure~\ref{Fig_Sc}b, since the high $T$ limit of $s_c$, $s_c^*$, depends strongly on monomer structure, chain stiffness, and cohesive interaction strength.~\cite{2008_ACP_137_125, 2014_Mac_47_6990} It would then appear that $s_c^*$, which can be estimated from simulation,~\cite{2013_JCP_138_12A541} provides a measure of the configurational anharmonicity of the intermolecular interactions that might have useful applications in the determination of optimal intermolecular potentials for simulations of fluid properties. The GET allows for the calculation of $s_c^*$ as a function of molecular and thermodynamic parameters. The results in the inset of Figure~\ref{Fig_Sc}b indicate that $s_c^*$ has significance for understanding thermodynamic scaling and deserves further systematic investigation.

We next consider how thermodynamic scaling impacts the segmental structural relaxation time $\tau_{\alpha}$ in the GET to gain insight about the consistency of the exponent $\gamma$ in both the Arrhenius and non-Arrhenius relaxation regimes. The GET involves no free parameters other than molecular parameters and thermodynamic variables, such as $P$ and $T$, but the model does involve some assumptions that we have discussed elsewhere.~\cite{2008_ACP_137_125, 2020_Mac_53_7239, 2021_Mac_XX_XXXX} Figure~\ref{Fig_Tau_GET}a shows our predictions of $\tau_{\alpha}$ over a range of $T$ for a range of fixed $P$, and Figure~\ref{Fig_Tau_GET}b shows the corresponding reduced variable plot indicating that all our data can be reduced to a universal scaling curve in the entire range of glass formation from $T_A$ to a $T$ range below the glass transition temperature $T_{\mathrm{g}}$, where equilibrium measurements are normally not even possible. We view the reduction Figure~\ref{Fig_Tau_GET}b as being truly remarkable. Importantly, the reduced density-temperature variable is the same in the $T$ regime below $T_A$ as in the Arrhenius regime so that our arguments above for the calculation of $\gamma_M$ apply to the entire liquid state regime.

\subsection{Thermodynamic Scaling based on Molecular Dynamics Simulation}

We further examine the remarkable thermodynamic scaling through a consideration of the extent of collective motion in GF liquids, as quantified by the average length $L$ of stringlike cooperative motion observed generally in simulations of GF liquids,~\cite{1998_PRL_80_2338, 2013_JCP_138_12A541, 2014_JCP_140_204509, 2006_PRL_97_045502, 2014_NatCommun_5_4163, 2015_JCP_142_234907, 2019_JPCB_123_5935, 2015_PNAS_112_2966, 2016_Mac_49_8355, 2016_MacroLett_5_1375, 2017_Mac_50_2585, 2017_SoftMatter_13_1190, 2019_JCP_150_101103, 2015_JCP_142_164506, 2013_ACP_152_519, 2013_ACP_152_519, 2018_JCP_149_161101, 2013_SoftMatter_9_241, 2011_PRL_106_115702, 2020_JCP_152_054904, 2020_Mac_53_4796, 2020_Mac_53_6828, 2020_Mac_53_9678} which has been found to play a key role in determining the activation energy for relaxation and which is closely related to the configurational entropy in the GET.~\cite{2008_ACP_137_125} In particular, many past works have shown that $L / L_A$, where $L_A$ is a residual collective motion occurring at $T_A$ and is something not envisioned in the original AG model,~\cite{1965_JCP_43_139} corresponds to the AG counterpart, $s_c^*/ s_c$. The determinations of $T_A$ and $L$ from simulations can be found in our previous works.~\cite{2020_Mac_53_6828, 2020_Mac_53_9678} Notably, AG also did not consider constant $V$ versus constant $P$ conditions, which can greatly influence how the model is `interpreted'.~\cite{2008_ACP_137_125} Note also that even apart from the difference between the configurational entropy and the excess entropy, the usual experimental choice, molar excess entropy, instead of its entropy density counterpart, certainly does not approach a plateau at high $T$ for fluids under constant $P$ conditions.~\cite{2008_ACP_137_125} We stress again that only the configurational entropy density ratio exhibits thermodynamic scaling in the GET.

\begin{figure*}[htb!]
	\centering
	\includegraphics[angle=0,width=0.975\textwidth]{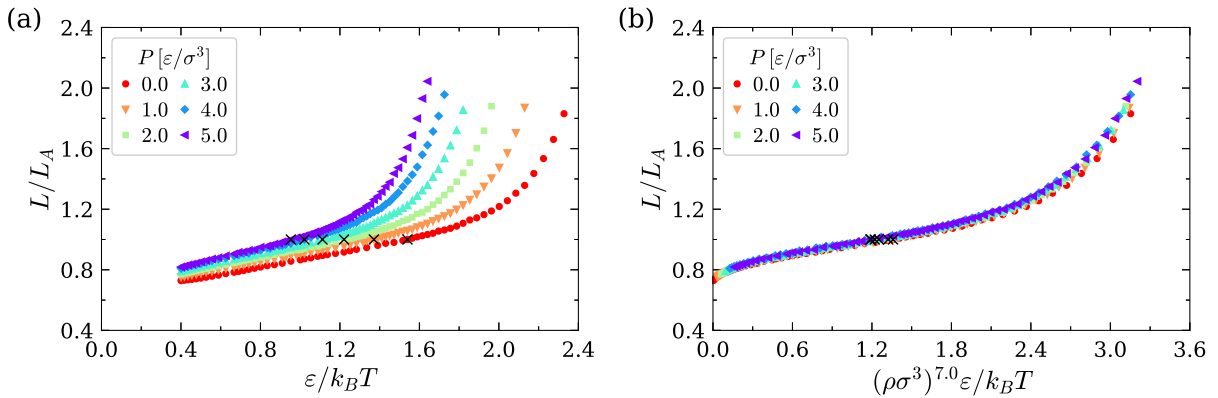}
	\caption{\label{Fig_LString}Simulation test of thermodynamic scaling of the extent of stringlike cooperative motion. (a) Normalized average string length $L/L_A$ versus inverse temperature $\varepsilon/k_BT$ for a range of $P$. The cross symbols indicate the positions of the onset temperature $T_A$ of glass formation. (b) $L/L_A$ versus $(\rho \sigma^3)^{\gamma}\varepsilon/k_BT$ with $\gamma =7.0$.}
\end{figure*}

\begin{figure*}[htb!]
	\centering
	\includegraphics[angle=0,width=0.975\textwidth]{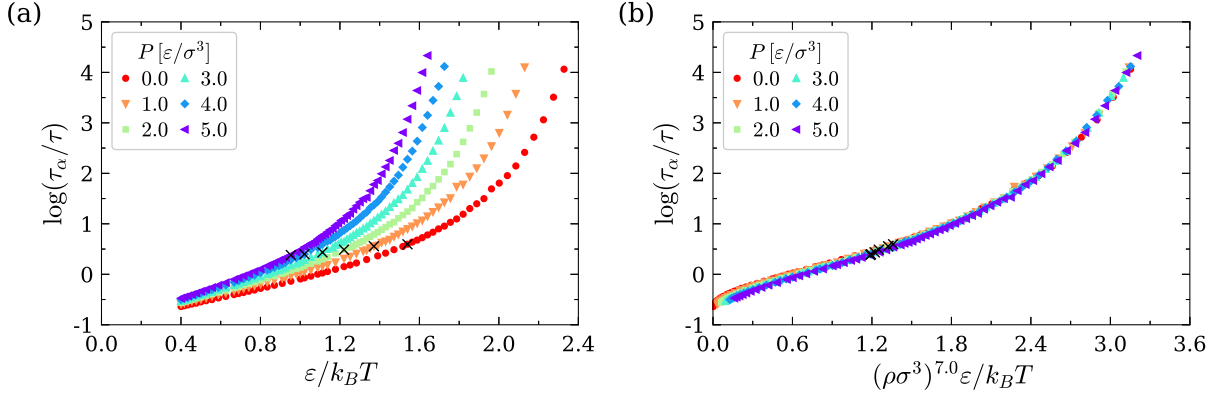}
	\caption{\label{Fig_Tau_MD}Simulation test of thermodynamic scaling of the structural relaxation time. (a) Logarithm of the structural relaxation time $\tau_{\alpha}$ versus inverse temperature $\varepsilon/k_BT$ for a range of $P$. The cross symbols indicate the positions of the onset temperature $T_A$ of glass formation. (b) $\log \tau_{\alpha}$ versus $(\rho \sigma^3)^{\gamma}\varepsilon/k_BT$ with $\gamma =7.0$.}
\end{figure*}

Accordingly, we plot $L / L_A$ versus $\varepsilon/k_BT$ in Figure~\ref{Fig_LString}, based on the simulations of a model of flexible polymer melts over a wide range of $T$ for a range of fixed $P$ that we have described in a previous work.~\cite{2020_Mac_53_6828} As in Figure~\ref{Fig_Sc}, we first show our results for $L / L_A$ for a range of fixed $P$ in Figure~\ref{Fig_LString}a and then in the thermodynamic scaling form in Figure~\ref{Fig_LString}b. It is evident that $L / L_A$ in our simulated polymer melt exhibits thermodynamic scaling as in our GET analysis of $s_c^*/ s_c$ where the scaling exponent $\gamma$ is again near $7.0$, a value consistent with typical values of $\gamma_M$ from EOS studies.~\cite{1967_JAP_38_4234, 2008_PRB_78_134201} The observation of thermodynamic scaling is also consistent with the `identification' of the strings with the abstract `cooperatively rearranging regions' hypothesized by AG.~\cite{1965_JCP_43_139}

Again mirroring our analysis based on the GET, Figure~\ref{Fig_Tau_MD} shows our simulation estimates of $\tau_{\alpha}$ from the intermediate scattering function~\cite{2020_Mac_53_6828} over a range of $T$ for a range of fixed $P$, where we first show $\log \tau_{\alpha}$ versus $\varepsilon/k_BT$ in Figure~\ref{Fig_Tau_MD}a and the corresponding reduction based on thermodynamic scaling in Figure~\ref{Fig_Tau_MD}b. The thermodynamic scaling reduction of $\tau_{\alpha}$ is excellent, as found for the viscosity, diffusion coefficient, and other dynamical properties for diverse GF liquids.~\cite{2005_RPP_68_1405} We can’t help but marvel about this remarkable phenomenon and its mathematical and physical origin. In particular, it is not clear why $s_c^*/ s_c$ should exhibit the thermodynamic scaling property, given that the GET results are completely insensitive to the form of the intermolecular potential. Could there be some effective many-body potential responsible for this remarkable scaling? It is notable that the original continuum mechanics type of arguments by Murnaghan~\cite{1944_PNAS_30_244, Book_Murnaghan} to deduce his highly successful EOS for condensed matter only required the $P$ dependence of the bulk modulus $B$ to be nearly linear over a wide range of $P$. This result, in conjunction with the theoretical derivation of the ME by Gilvarry,~\cite{1956_PR_102_331, 1957_JAP_28_1253} based on an extension of the Debye theory to include anharmonicity and the quantification of anharmonicity suggested by Gr{\"u}neisen,~\cite{1955_JCP_23_1925} indicates that anharmonic interactions are apparently the key to understanding thermodynamic scaling and by extension variability in collective motion and fragility of GF liquids, as these interactions are critical for understanding the existence of thermal expansion of solids and $T$-dependent moduli~\cite{1958_PM_8_1381} and the thermal conductivity of non-metallic solid materials.~\cite{2018_MTP_4_50, 2020_APL_117_123901}

We note that the work of Sastry and coworkers\cite{2013_EPJE_36_141} indicates that the AG model exhibits thermodynamic scaling for the structural relaxation time in the Kob-Anderson model if the density scaling of the high temperature activation energy parameter of the AG model (designated as $A$) is considered. This is broadly consistent with our results, along with the experimental observations of Masiewicz et al.~\cite{2015_SciRep_5_13998} These findings are purely empirical, but they support the origin of thermodynamic scaling as being related to the scaling of $s_c$.

\subsection{Thermodynamic Scaling as Filter for Assessing Theories of Glass Formation}

Dyre and coworkers~\cite{2009_JCP_131_234504} have argued that thermodynamic scaling, apart from its theoretical and practical significance, has the additional value as offering a `filter' for assessing theories of glass formation. Even though one may argue about why thermodynamic scaling exists, theories that do not give rise to this property should be rejected. We discuss the power of this way of thinking by considering some alternative models than the AG model~\cite{1965_JCP_43_139} in relation to thermodynamic scaling. In particular, there are alternative models of the dynamics of GF liquids that emphasize, for example, the dependence of the structural relaxation time and diffusion coefficient on the static structure factor in the long wavelength limit, $S(0)$,~\cite{2004_JCP_121_1984, 2004_JCP_121_2001, 2010_ARCMP_1_277, 2014_JCP_140_194506, 2014_JCP_140_194507, 2015_Mac_48_1901, 2016_Mac_49_9655} and the mean square displacement on a caging timescale on the order of a picosecond, i.e., the so-called `Debye-Waller parameter', $\langle u^2 \rangle$.~\cite{2008_NaturePhys_4_42, 2012_SoftMatter_8_11455} It is not obvious on the surface whether these models are related to the AG model, and whether the central quantities in these models should exhibit thermodynamic scaling. 

\begin{figure*}[htb!]
	\centering
	\includegraphics[angle=0,width=0.975\textwidth]{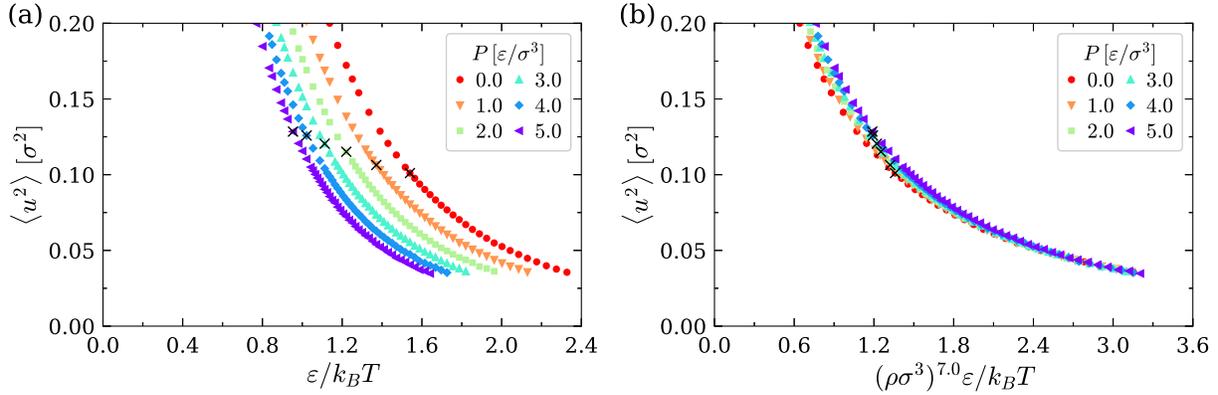}
	\caption{\label{Fig_DWF}Simulation test of thermodynamic scaling of the Debye-Waller parameter. (a) Debye-Waller parameter $\langle u^2 \rangle$ versus inverse temperature $\varepsilon/k_BT$ for a range of $P$. The cross symbols indicate the positions of the onset temperature $T_A$ of glass formation. (b) $\langle u^2 \rangle$ versus $(\rho \sigma^3)^{\gamma}\varepsilon/k_BT$ with $\gamma =7.0$. The result in panel (b) shows a reasonably successful, but not perfect, transformation of $\langle u^2 \rangle$ to a reduced variable description consistent with the thermodynamic scaling exhibited by $\tau_{\alpha}$.}
\end{figure*}

\begin{figure*}[htb!]
	\centering
	\includegraphics[angle=0,width=0.975\textwidth]{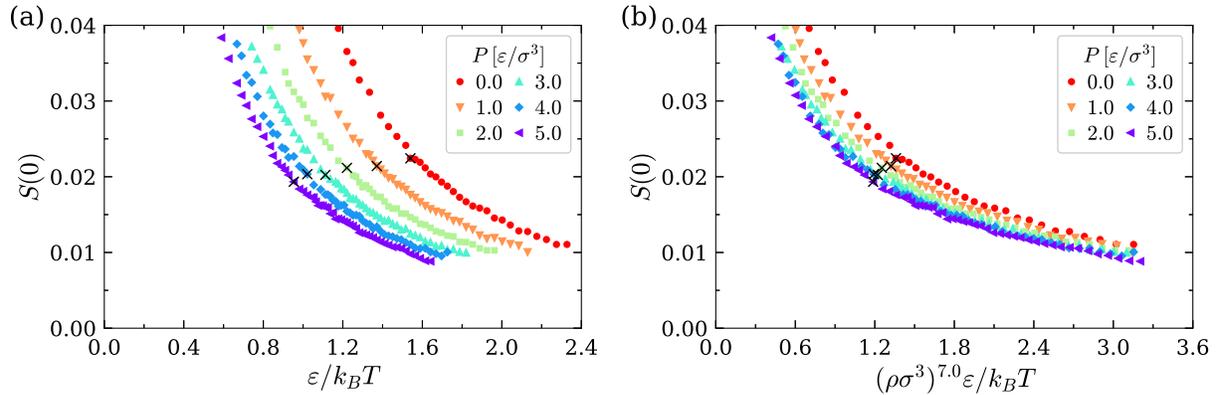}
	\caption{\label{Fig_S0}Simulation test of thermodynamic scaling of the long wavelength limit of the static structure factor. (a) Long wavelength limit $S(0)$ of the static structure factor versus inverse temperature $\varepsilon/k_BT$ for a range of $P$. The cross symbols indicate the positions of the onset temperature $T_A$ of glass formation. (b) $S(0)$ versus $(\rho \sigma^3)^{\gamma}\varepsilon/k_BT$ with $\gamma =7.0$. The result in (b) shows that a transformation to a reduced variable description based on the thermodynamic scaling exhibited by $\tau_{\alpha}$ is not successful for $S(0)$.}
\end{figure*}

Figures~\ref{Fig_DWF} and~\ref{Fig_S0} show our attempt at finding a reduced scaling for $\langle u^2 \rangle$ and $S(0)$ estimated from our simulations. In Figure~\ref{Fig_DWF}, we find that thermodynamic scaling provides a reasonable description of our simulation observations of $\langle u^2 \rangle$, although the thermodynamic scaling of $\langle u^2 \rangle$ is apparently not as perfect as in the case of $s_c/ s_c^*$. Further investigations are required to determine whether the small deviations from thermodynamic scaling arise from numerical uncertainties or other factors not yet identified.

The deviation from thermodynamic scaling is well beyond numerical uncertainties in the case of $S(0)$ in Figure~\ref{Fig_S0}. $S(0)$ is proportional to the isothermal compressibility of the fluid, so one might have thought that such a quantity would exactly exhibit the pressure-potential energy correlations emphasized by Dyre and coworkers~\cite{2009_JCP_131_234504} as a possible origin of thermodynamic scaling in many real liquids. While the estimates of $S(0)$ based on $S(q)$ may depend on the system size, we have confirmed that varying the system size over a reasonable range does not change our results for $S(0)$ in any substantial way. The inordinately large-scale simulations of Chremos~\cite{2020_JCP_153_054902} indicate that our estimates of $S(0)$ in the present work are reliable. Apparently, the system cannot be too small to make reliable estimates of $S(0)$. Our simulations and the work of Chremos show that the lowest wavenumber $q$ should be $\sim 0.5\sigma^{-1}$ in the present polymer model, which corresponds to a total bead number of $\sim 2000$ for a typical number density of $\rho =1$. Moreover, it should be noted that the estimates of $S(0)$ by first calculating $\kappa_T$ and $\rho$ by the usual sum rule for $S(0)$ may yield somewhat different results, where the difference should be larger for lower $T$. There are long-range correlations developing in the system that cause deviations from the sum rule relating $\kappa_T$, $\rho$, and $k_BT$ to $S(0)$. Since long-range correlations can be expected in GF liquids, this matter should be investigated further in the future.

The simple fact that $S(0)$ clearly does not exhibit thermodynamic scaling means that we may reject the mode-coupling theory~\cite{Book_Gotze} and the nonlinear Langevin equation theory~\cite{2004_JCP_121_1984, 2004_JCP_121_2001, 2010_ARCMP_1_277} and its recent extension, the elastically cooperative nonlinear Langevin theory~\cite{2014_JCP_140_194506, 2014_JCP_140_194507, 2015_Mac_48_1901, 2016_Mac_49_9655} as acceptable models of GF liquids, based on the `filter criterion' of Dyre and coworkers~\cite{2009_JCP_131_234504}. Berthier and Tarjus~\cite{2010_PRE_82_031502} have come to the essentially same conclusion regarding the physically inadequacy of the mode-coupling theory and its variants because this class of `generalized free volume' theories presume that the pair correlation function $g(r)$, or its integral $S(0)$, determines the dynamics of fluids. In physical terms, our analysis means that `structure', based on $S(0)$ or properties derived from $g(r)$, does not by itself determine the dynamics of liquids.

Dyre and coworkers~\cite{2019_JCP_151_204502} have clarified the origin of the `breakdown' of thermodynamic scaling of $S(0)$ in condensed fluids through the analysis of many of the virial-related thermodynamic properties of materials composed of atomic species interacting through the ubiquitous LJ potential. As well-known,~\cite{1972_JCP_57_1259, 1979_JCP_70_1837} thermodynamic scaling is exactly recovered if the attractive contribution to the LJ potential is formally assumed to vanish, but this power-law density-temperature scaling no longer holds when the attractive component to the LJ potential is included.~\cite{2019_JCP_151_204502} An attractive interaction is essential for the stability of condensed matter at constant $P$ so we may rather generally expect virial-related properties, such as $S(0)$, to not exhibit thermodynamic scaling, a result that accords with our analytic results based on the GET and our simulations of polymeric GF liquids discussed below.

Of course, the loss of `thermodynamic scaling' does not imply that a more general density-temperature scaling does not exist. Dyre and coworkers~\cite{2013_PRE_88_042139, 2014_JPCB_118_10007} have shown how to derive generalized density-temperature scaling relations from the same type of potential-virial correlations as found in the case of a purely repulsive power-law interatomic potential, and they have discussed the significance of these generalized density-temperature scaling relations, when they exist, for the dynamics of model condensed materials. In particular, they have devised a classification scheme of liquids (Roskilde simplicity classification of liquids based on `isomorphs') based on these extended density-temperature scaling relationships.~\cite{2013_PRE_88_042139, 2014_JPCB_118_10007}

Recent computational studies of a range of GF liquids, including polymeric,~\cite{2014_JCP_141_054904, 2015_JCP_143_194503, 2016_JCP_145_234904} ionic~\cite{2011_JCP_134_144507, 2011_JCP_135_164510} and certain metallic GF liquids,~\cite{2016_JCP_145_104503} have shown that the virial-potential scaling that is central to the existence of generalized density-temperature scaling does not hold, which minimally seems to imply that this classification scheme is of limited value for polymeric GF liquids, which are our main interest. Nonetheless, the work of Dyre and coworkers~\cite{2019_JCP_151_204502} provides significant insight into how attractive interactions `break the scaling symmetry' responsible for the simple power-law scaling of the compressibility factor, and other thermodynamic properties related to the pair correlation function. This finding has far reaching implications for the theories of the dynamics of GF liquids, such as the conventional mode-coupling theory of glass formation,~\cite{Book_Gotze} that predict that a relation between the structural relaxation time and equilibrium `structural parameters' derived from $g(r)$, such as $S(0)$, and the isothermal compressibility, $\kappa_T$. We note that a lack of consistent density-temperature scaling also arises in the peak height of the $4$-point density function, a higher-order density correlation function than $S(0)$, which is often considered to be a measure of `dynamic heterogeneity' in GF liquids.~\cite{2013_JPCL_4_4273}
	
Recently, there have been efforts to address the problems in existing mode-coupling type formulations of the dynamics of GF liquids just noted. Dell and Schweizer~\cite{2015_PRL_115_205702} recognized the general failure of mode-coupling approaches in which collective interparticle interactions are modeled by $S(0)$, and they addressed this fundamental problem in an extended mode-coupling approach, along with a series of approximations, to account for the effect of attractive interparticle interactions. Interestingly, their extended mode-coupling theory framework can apparently account for many of the shortcomings identified in earlier simulations of model atomic GF liquids based on the assumption that interparticle interactions are dominated by pairwise repulsive interactions alone. Perhaps most importantly from the standpoint of the present paper, Dell and Schweizer~\cite{2015_PRL_115_205702} found that their extended theory was consistent with thermodynamic scaling when applied to model atomic GF liquids studied previously. This promising extension of mode-coupling theory, along with novel additional elements included by Dell and Schweizer,~\cite{2015_PRL_115_205702} requires further validation and has not been applied to the much more complicated case of polymeric GF liquids. At this stage of development, we note that the structural relaxation time of even these toy atomic fluids is not simply a function of $S(0)$ in the proposed modified mode-coupling theory so the absence of thermodynamic scaling for $S(0)$ cannot be used to exclude the new model of Dell and Schweizer.~\cite{2015_PRL_115_205702} It will be interesting to see how this improved mode-coupling theory performs in the future for more complex GF liquids.

We shall see below that we may recover a consistent thermodynamic scaling by subjecting the isothermal compressibility to a density dependent transformation that relates this property to the configurational entropy, allowing us to obtain a reduced isothermal compressibility that satisfies thermodynamic scaling. These relations give us hope of a thermodynamic relation between the dynamics and thermodynamics of GF liquids built around the configurational entropy density and properties related to the configurational entropy density.

We finally point out that the occurrence of thermodynamic scaling is not itself a sufficient criterion for the validity of a model of glass formation. Fluids that exhibit strong correlations between the virial and potential energy fluctuations from the existence of a power-law intermolecular potential exhibit thermodynamic scaling, but molecular fluids do not normally exhibit strongly correlating dynamics because the presence of chemical bonds tends to lead to the loss of these correlations. This situation would seem to preclude the use of atomic GF liquids as model systems for describing the dynamics of molecular GF fluids. Of course, purely density-based `free volume' models of glass formation can be ruled out because such models are inherently inconsistent with thermodynamic scaling. Dyre and coworkers~\cite{2009_JCP_131_234504} are indeed correct in indicating that thermodynamic scaling offers an important criterion for checking the validity of models of glass formation. We view the violation of thermodynamic scaling for $S(0)$ and other properties related to the fluid pair correlation function $g(r)$ as providing further evidence in favor of the interpretation of thermodynamic scaling originally proposed by Casalini and Roland.~\cite{2007_JPCM_19_205118}

\subsection{Relationship between Configurational Entropy and Isothermal Compressibility}

Although the GET is helpful in understanding the origin of thermodynamic scaling as being fully consistent with this model of the dynamics of GF liquids, the inherent difficulty, if not plain impossibility, of estimating the configurational entropy density $s_c$ in measurements is certainly a downside of the entropy model of glass formation. Of course, there is no real problem if we express the thermodynamic scaling from the GET in terms of temperature and molecular parameters, but this situation makes us wonder if there are any other thermodynamic properties that exhibit thermodynamic scaling, which are more readily accessible experimentally and might serve as a surrogate for $s_c$ in the GET. It is certainly helpful that the GET allows us to calculate any thermodynamic property that we can think of to examine the above possibility. The fragility parameter $D$ of the VFT equation or the steepness parameter $m$, quantifying the strength of the $T$ dependence of the structural relaxation and diffusion at constant pressure or volume, may be calculated in terms of only the molecular and thermodynamic parameters required to specify the thermodynamic properties of the material. It is clearly of some advantage that we have a fully predictive theory of all these quantities in our quest for an alternative to $s_c$.

As others had done before,~\cite{2011_JCP_135_164510} we have found that the Debye-Waller parameter $\langle u^2 \rangle$ exhibits thermodynamic scaling to a good approximation, which gives us hope that some other thermodynamic properties might obey thermodynamic scaling. This observation can further be taken as a major clue of where we should look for such a property that encodes similar information as $s_c$.

We have long viewed $\langle u^2 \rangle$ as corresponding to a kind of `dynamical free volume' that is related to the local compressibility rather than a static geometrical size of cavities on a scale set by the particles by density fluctuations as in traditional `free volume' models emphasizing density and the `structure' of the fluid. $\langle u^2 \rangle$ largely reflects the thermal energy of the molecules and the same can be said to be true of the isothermal compressibility $\kappa_T$ of the material, which is the inverse of the bulk modulus, $B$. $\kappa_T$ is clearly accessible by mechanical or sound propagation measurements or through measurements of the structure factor at long scattering wavelengths, $S(0)$, in conjunction with estimates of the density through a well-known thermodynamic relation, $\kappa_T = \rho k_BT/S(0)$. $\kappa_T$ is also an attractive quantity to consider for developing a description of the dynamics of GF liquids because, as we have discussed in Section~\ref{Sec_Intro}, this property is central to defining the EOS of condensed materials and its derivatives clearly encode information about the anharmonicity of the interparticle interactions in the condensed state. While it is clear from a direct analysis based on the GET, and validating coarse-grained MD simulations, that $\kappa_T$ does not normally exhibit thermodynamic scaling in molecular GF liquids, we might then consider if any obvious transformation of $\kappa_T$ might give rise to the definition of a reduced compressibility which has the thermodynamic scaling property and which has some relationship to $s_c/ s_c^*$. We next show that this is indeed the case.

We start by noting that the $T$ variation of $\langle u^2 \rangle$ varies in a way that changes as one approaches various temperatures that are characteristic of glass formation, and thus, our transformation of $\kappa_T$ should show some recognition of these characteristic temperatures. For example, we have repeatedly observed over the admittedly limited $T$ range accessible to equilibrium simulations that $\langle u^2 \rangle$ at low $T$ scales linearly with $T$ and apparently extrapolates to zero at a temperature consistent with the VFT temperature $T_0$, deduced from the structural relaxation time.~\cite{2002_PRL_89_125501, 2009_PNAS_106_7735, 2016_JSM_054048, 2018_JCP_148_104508, 2020_JCP_153_124508} The word `extrapolation' is emphasized here as we do not literally believe that $\langle u^2 \rangle$ vanishes at $T_0$ any more than we think that $s_c$ actually vanishes at its corresponding Kauzmann temperature $T_K$, which likewise relies on an extrapolation from the $T$ regime above $T_{\mathrm{g}}$. In the GET, $T_K$ is basically equivalent to $T_0$ because the VFT functional form is predicted to hold rather well over the $T$ range between $T_{\mathrm{g}}$ and the crossover temperature $T_c$, which is typically about $(1.2 - 1.3) T_{\mathrm{g}}$, depending on chain length, chain stiffness, cohesive interaction strength, monomer structure, pressure, etc. 

The thermodynamic scaling of $\langle u^2 \rangle$ at low $T$ suggests that we consider the \textit{relative} isothermal compressibility, $\Delta \kappa_T \equiv \kappa_T - \kappa_{T,0}$ with $\kappa_{T,0}$ being the value of $\kappa_T$ at $T_0$, which mathematically acknowledges that there are `special' temperatures involved in glass formation and which by construction vanishes at $T_0$. It is important to note that $T_0$ can be measured experimentally and $\kappa_{T,0}$ can be computed precisely in the GET, so we remain in complete theoretical control and are discussing quantities that can be clearly measured. Note that $\kappa_{T,0}$ depends on molecular parameters and pressure so that $\Delta \kappa_T$ may have very different scaling properties than $\kappa_T$ when both $T$ and $V$ are varied.

Since $\Delta \kappa_T$ has the units of a reciprocal energy density, we are still comparing thermodynamic `apples and oranges', so to speak, if we attempt to compare to the configurational entropy density ratio $s_c^*/ s_c$. We then consider some sort of `obvious' reduced variable that might be interrelated to $s_c^*/ s_c$, which is the central quantity in the GET controlling the $T$ dependence of the activation energy for $\tau_{\alpha}$. We note here that $s_c^*$ is highly dependent on molecular parameters, so reducing $s_c$ by $s_c^*$ is clearly the key to the existence of the thermodynamic scaling of $s_c^*/ s_c$. Given our definition of $s_c^*/ s_c$, it seems natural to define a reduced $s_c$ that vanishes at the onset temperature $T_A$ of glass formation, another characteristic temperature that may be calculated precisely in the GET and estimated experimentally. In particular, we introduce a reduced relative configurational entropy density $\delta s_c$,
\begin{equation}
	\label{Eq_ScReduced}
	\delta s_c \equiv s_c^*/ s_c - 1 = (s_c^* - s_c)/s_c,
\end{equation} 
which vanishes by construction at $T_A$. This definition may be recognized as being analogous to the type of reduced variable introduced in critical phenomena in the development of a universal reduced description of thermodynamic and dynamic properties where $s_c$ formally replaces $T$. Note that $\delta s_c$ naturally arises in the formal separation of the activation free energy of the GET/AG model into uncooperative and cooperative components,~\cite{2008_ACP_137_125} i.e., $\Delta G(T) (s_c^*/ s_c) = \Delta G_o + \Delta G_o \delta s_c$.

\begin{figure*}[htb!]
	\centering
	\includegraphics[angle=0,width=0.975\textwidth]{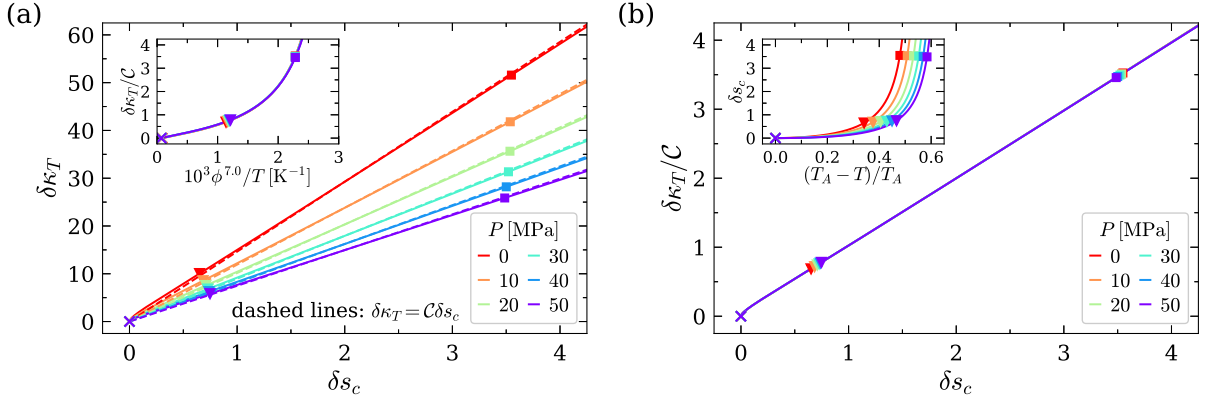}
	\caption{\label{Fig_KappaTSc}Relationship between the isothermal compressibility and configurational entropy density in the generalized entropy theory. (a) Reduced isothermal compressibility $\delta \kappa_T$ versus reduced configurational entropy density $\delta s_c$ for a range of $P$. The cross, triangle, and square symbols indicate the positions of the onset $T_A$, crossover $T_c$, and glass transition temperatures $T_{\mathrm{g}}$ of glass formation, respectively. Dashed lines are fits to the equation, $\delta \kappa_T = \mathcal{C} \delta s_c$, where $\mathcal{C}$ depends on $P$. (b) $\delta \kappa_T / \mathcal{C}$ versus $\delta s_c$ for a range of $P$. The inset in panel (a) displays the thermodynamic scaling of the ratio $\delta \kappa_T / \mathcal{C}$. Symbols in panel (b) have the same meaning as in panel (a). While there is no strong `kink' in $\delta \kappa_T$, as one would expect for the presence of a phase transition, there are small changes in slope around $T_A$ and $T_c$.}
\end{figure*}

Now if we define a corresponding dimensionless reduced variable for $\Delta \kappa_T$, then we will be in a position to directly compare $\kappa_T$ and $s_c$ in an appropriate form that recognizes the existence of characteristic temperatures at which the thermodynamics and dynamics of GF liquids change in a significant fashion. We then define a dimensionless isothermal compressibility as,
\begin{equation}
	\label{Eq_KappaTReduced}
	\delta \kappa_T \equiv (\kappa_{T,A} - \kappa_T) / (\kappa_T - \kappa_{T,0}),
\end{equation} 
where $\kappa_{T,A}$ is the value of $\kappa_T$ at $T_A$. This dimensionless quantity vanishes at $T_A$, as in the case of $\delta s_c$. However, it should be noted that our reduced variables are not well-defined as $T$ approaches $T_0$, but this extremely low $T$ regime does not interest us practically because the GET has limited interest below $T_{\mathrm{g}}$ where equilibrium measurements on GF liquids are normally impossible. Accordingly, we might expect $T_{\mathrm{g}}$ to somehow come into our reduced variable description of $\delta \kappa_T$, since $T_{\mathrm{g}}$ is clearly an important characteristic temperature of glass formation. We show the comparison of $\delta \kappa_T$ to $\delta s_c$ for different $P$ in Figure~\ref{Fig_KappaTSc}a. These quantities are proportional to a high degree of approximation, if not equality, and we may fix the prefactor by invoking a condition that the reduced variable slopes should coincide by introducing $T_{\mathrm{g}}$ as a reference point. In particular, we find the reduced variable relation,
\begin{equation}
	\label{Eq_KappaTSc}
	\delta \kappa_T = \mathcal{C} \delta s_c,
\end{equation} 
where the $P$-dependent proportionality factor may be given by $\mathcal{C} = \delta \kappa_T(T_{\mathrm{g}}) / \delta s_c(T_{\mathrm{g}})$ by taking $T_{\mathrm{g}}$ as a reference point. We plot $\delta \kappa_T / \mathcal{C}$ in Figure~\ref{Fig_KappaTSc}b, where we see that $\delta \kappa_T / \mathcal{C}$ seems to scale perfectly linearly with $\delta s_c$. The inset of Figure~\ref{Fig_KappaTSc}a indicates that $\delta \kappa_T / \mathcal{C}$ obeys thermodynamic scaling. It goes without saying that $\delta s_c$ obeys thermodynamic scaling, which is demanded by mathematical consistency and was checked explicitly. We now have a thermodynamic property exhibiting thermodynamic scaling defined in terms of experimentally measurable properties. As a separate matter, the inset of Figure~\ref{Fig_KappaTSc}b shows the $T$ dependence of $\delta s_c$, which is nearly quadratic in the reduced temperature in the high $T$ regime near $T_A$, i.e., $\delta s_c = C_s [(T_A - T)/T_A]^2$ with $C_s$ being a material specific constant,~\cite{2008_ACP_137_125} but in the low $T$ regime below $T_c$, the $T$ dependence of $s_c$ changes to a variation in which the product of $s_c$ and $T$ is linear, i.e., $Ts_c = K [(T - T_0)/T_0]$, where $K$ is a constant defining the thermodynamic fragility of glass formation.~\cite{2008_ACP_137_125} The $T$ variation of $s_c$ predicted by the GET has recently been reviewed in ref~\citenum{2021_Mac_XX_XXXX}.

Quite apart from achieving our practically significant goal of transforming our entropy theory to a model centered on the thermodynamically `natural' variable of the isothermal compressibility, we may as well ask why $\kappa_T$ itself does not exhibit thermodynamic scaling. Why do we need such a transformation to restore the thermodynamic scaling property? As noted before and recognized when developing the transformation, the situation here is very similar to that arising in the second-order phase transition at which a special critical temperature arises so that thermodynamic and dynamic properties of the system do not depend on $T$ alone. Specifically, in the $O(n)$ class of spin models describing the Ising model ($n = 1$), X-Y model ($n = 2$), Heisenberg model ($n = 3$), …, spherical model ($n \rightarrow \infty$), the thermodynamic and dynamic properties become universal functions of a reduced temperature, $T = (T - T_{cr}) / T$, with a non-universal prefactor near the precisely defined critical temperature $T_{cr}$, where both the prefactor and $T_{cr}$ depend on the interaction strength, material structure, order parameter dimension $n$, spatial dimension, etc. However, thermodynamic scaling is lost when the dimension is reduced below the lower critical dimension at which $T_{cr}$ becomes zero so that the properties become only a function of $T$, and no finite critical temperature and thus nontrivial reduced critical temperature exists. This situation is realized, for example, in the Ising model in one dimension.~\cite{Book_Baxter} It would appear that the breaking of thermodynamic scaling in $\kappa_T$, and other properties for which virial-potential scaling might have been anticipated, can be traced to a thermodynamic transition phenomenon underlying glass formation that is characterized by well-defined characteristic temperatures defined by the equilibrium thermodynamics of the material. We have provided evidence for this scenario in a previous work,~\cite{2006_JCP_125_144907} where we have emphasized that the transition involves a particle supramolecular assembly transition responsible for the emergence of dynamic heterogeneity in GF liquids, which is a type of rounded transitions. It is characteristic of such self-assembly transitions that they have both onset and ending transition temperatures analogous to $T_A$ and $T_{\mathrm{g}}$ and a crossover transition in the middle analogous to $T_c$ so that the need for multiple characteristic temperatures rather than a single $T_c$ is consistent with the proposed picture of the thermodynamic nature of glass formation and conception of the origin of dynamic heterogeneity in such transitions as a generic and dynamic phenomenon in condensed matter. Importantly, this theoretical framework, which is embodied by the string model of glass formation,~\cite{2013_JCP_138_12A541, 2006_PRL_97_045502, 2014_NatCommun_5_4163, 2014_JCP_140_204509, 2015_PNAS_112_2966, 2015_JCP_142_234907, 2019_JPCB_123_5935, 2016_Mac_49_8355, 2016_MacroLett_5_1375, 2017_Mac_50_2585, 2019_JPCB_123_5935, 2020_JCP_152_054904, 2020_Mac_53_4796, 2020_Mac_53_6828, 2020_Mac_53_9678} provides a framework for calculating and interpreting the dynamic properties of glass formation.

\begin{figure*}[htb!]
	\centering
	\includegraphics[angle=0,width=0.975\textwidth]{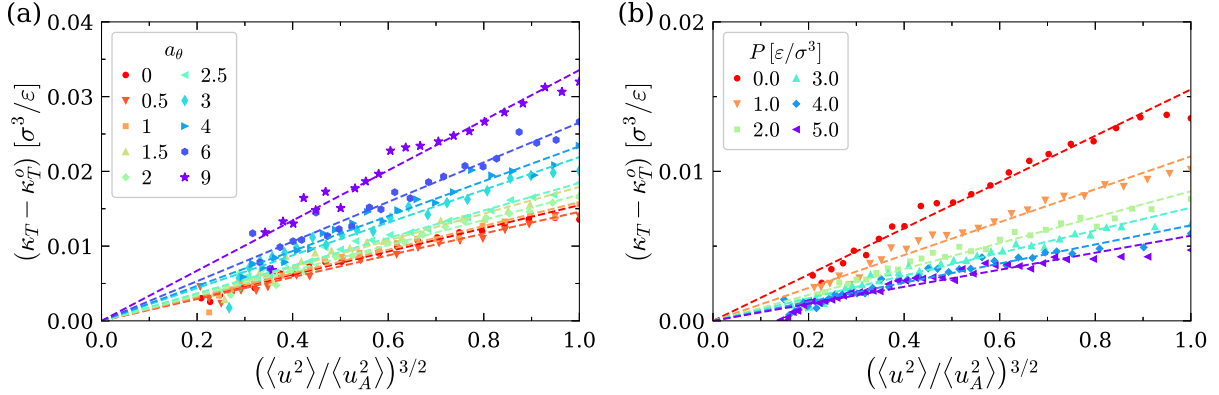}
	\caption{\label{Fig_KappaTDWF}Relationship between the isothermal compressibility and Debye-Waller parameter in simulations. Relative isothermal compressibility $(\kappa_T - \kappa_T^o)$ versus $\left( \langle u^2 \rangle / \langle u_A^2 \rangle \right)^{3/2}$ determined from simulations of polymer melts having variable chain stiffness~\cite{2020_Mac_53_4796} and pressure.~\cite{2020_Mac_53_6828} (a) Observations for a range of chain stiffness parameters $a_{\theta}$. (b) Observations for a range of $P$. Note that $\kappa_T^o$ is the extrapolated value of $\kappa_T$ at the temperature at which $\langle u^2 \rangle$ correspondingly extrapolates to zero. Dashed lines indicates descriptions according to eq~\ref{Eq_KappaTDWF}.}
\end{figure*}

We are now in a position to test our starting hypothesis that $\langle u^2 \rangle^{3/2}$ corresponds to a measure of the `local compressibility' by directly comparing $\kappa_T$ to $\langle u^2 \rangle^{3/2}$. This should also provide insight into the interrelation between the entropy theory of glass formation and the localization model,~\cite{2012_SoftMatter_8_11455, 2015_PNAS_112_2966, 2016_JSM_054048} which focuses on the seemingly different property, $\langle u^2 \rangle^{3/2}$. Historically, $\langle u^2 \rangle^{3/2}$ has been interpreted as a measure of dynamical free volume,~\cite{1972_JCP_57_1259, 1979_JCP_70_1837, 1986_JPC_90_6252, 1998_MP_95_289, 1997_PRE_56_5524, 2012_SoftMatter_8_11455} an interpretation that might illuminate the interrelation between the entropy and free volume approaches to the dynamics of GF liquids. The potential importance of this dynamical variant of free volume can be traced back to some of the earliest scientific studies of condensed matter, as illustrated below in our discussion of Sutherland's kinetic theory of solids.~\cite{1890_PM_30_318, 1891_PM_32_31, 1891_PM_32_215, 1891_PM_32_524} We point out that $\langle u^2 \rangle$ has also been interpreted in the theory of Leporini and coworkers~\cite{2008_NaturePhys_4_42} to be a measure of reciprocal rigidity or `compliance' in a rheological terminology. Hence, $\langle u^2 \rangle$ is clearly central to a number of models of glass formation where a remarkable range of interpretations have been given to this quantity.

Since we must determine $\langle u^2 \rangle$ in simulations, we consider its relation with $\kappa_T$ based on our simulation results. For generality, we also analyze the simulation results based on a coarse-grained polymer melt with variable chain stiffness parameter $a_{\theta}$. See ref~\citenum{2020_Mac_53_4796} for details about the polymer model. We first define $\kappa_T^o$ as the value of $\kappa_T$ at the temperature at which $\langle u^2 \rangle$ extrapolates to zero, a condition that corresponds to the VFT temperature in the localization model.~\cite{2012_SoftMatter_8_11455, 2015_PNAS_112_2966, 2016_JSM_054048} This correspondence has been confirmed in previous simulation studies of polymeric~\cite{2002_PRL_89_125501, 2018_JCP_148_104508} and metallic GF liquids~\cite{2016_JSM_054048, 2020_JCP_153_124508} and grain boundary mobility.~\cite{2009_PNAS_106_7735} However, we again emphasize that the vanishing of $\langle u^2 \rangle$ at a low $T$ should not be taken literally just as the divergence of $\tau_{\alpha}$ at the VFT temperature should not be taken literally. Taking the values of $\langle u^2 \rangle$ at $T_A$ as $\langle u_A^2 \rangle$, our analysis based on polymer melts with variable $a_{\theta}$ and $P$ in Figure~\ref{Fig_KappaTDWF} indicates that the relative isothermal compressibility, $(\kappa_T - \kappa_T^o)$, and the normalized quantity, $\left( \langle u^2 \rangle / \langle u_A^2 \rangle \right)^{3/2}$, are proportional to each other to a good approximation,
\begin{equation}
	\label{Eq_KappaTDWF}
	(\kappa_T - \kappa_T^o) = \mathcal{D} \left( \langle u^2 \rangle / \langle u_A^2 \rangle \right)^{3/2},
\end{equation} 
where the proportionality factor $\mathcal{D}$ may likewise be estimated by taking $T_{\mathrm{g}}$ as a reference point, as in eq~\ref{Eq_KappaTSc}. This finding is consistent with our hypothesis that $\langle u^2 \rangle^{3/2}$ can be interpreted as a local compressibility. The relative compressibility $(\kappa_T - \kappa_T^o)$ may be normalized to arrive at a dimensionless reduction of the data for $\kappa_T$, but we found that this procedure results in a plot with some scatter, given the large amount of data, so we do not consider such a reduction in the present exploratory study. Note that there is some curvature in some of the data for high $P$ and low $\langle u^2 \rangle$, which we do not currently understand, and we must admit that the best fitted exponent could be a value somewhat different than the value of $3/2$ assumed in Figure~\ref{Fig_KappaTDWF}. These are refinements beyond our main point that $\langle u^2 \rangle$ can qualitatively be interpreted as a measure of local fluid compressibility. We plan to study the interrelation between $\kappa_T$ and $\langle u^2 \rangle$ more thoroughly in the future work.

It is also suggested from a comparison of Figures~\ref{Fig_KappaTSc} and~\ref{Fig_KappaTDWF} that a relationship exists between $s_c$ and $\langle u^2 \rangle^{3/2}$, a relationship that was noticed, but not understood, in previous works.~\cite{2002_PRL_89_125501, 2015_PNAS_112_2966} These correspondences thus allow us to understand the occurrence of a common thermodynamic scaling for $\delta s_c$, $\delta \kappa_T$, and $\langle u^2 \rangle$, thereby providing some understanding of the interrelations between different models of glass formation that emphasize these thermodynamic properties in relation to dynamics.

We may obtain some insight into the critical value of $\delta \kappa_T$ from our previous study of a coarse-grained polymer melt under variable $P$ conditions,~\cite{2017_Mac_50_2585} which considers a dimensionless compressibility defined by the ratio of $S(0)$ to the value of the static structure factor at its first peak corresponding to an average intersegment distance, the so-called `hyperuniformity index', $h$. All the $h$ curves intersect at a common temperature that roughly equals the VFT temperature $T_0$ obtained from the structural relaxation time to within simulation uncertainty, where $h$ is near $10^{-3}$, the defining condition for effective hyperuniformity.~\cite{2016_Mac_49_8341, 2018_PR_745_1, 2018_PRL_121_258002, 2017_AP_529_1600342, 2020_JCP_153_054902} This intersection phenomenon of the isothermal compressibility curves is also apparent in ref~\citenum{2016_Mac_49_8341}, where we estimated $h$ when the cohesive energy parameter was varied at constant $P$ and then compared the results to those for variable $T$ conditions but at constant $V$. The intersection point appears to be about the same as in the variable pressure study, so it appears to be some sort of invariant. This critical condition has great potential relevance for defining an EOS for dense fluids that should have comparable significance for gases and liquids near their other critical point associated with the emergence of a hyperuniform state.~\cite{2018_PR_745_1}

\subsection{\label{Sec_Anharmonicity}Quantification of Anharmonicity in the Non-Arrhenius Relaxation Regime}

\begin{figure*}[htb!]
	\centering
	\includegraphics[angle=0,width=0.975\textwidth]{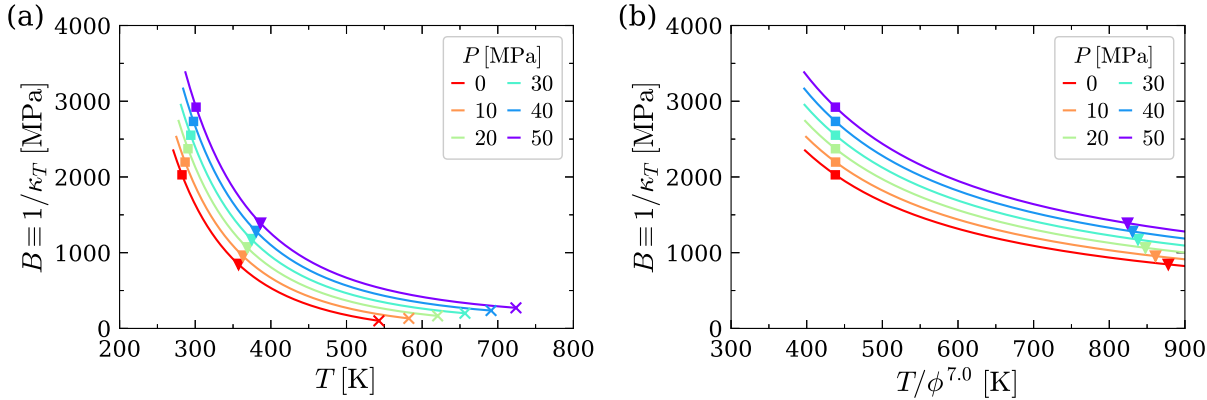}
	\caption{\label{Fig_BT_GET}Generalized entropy theory test of thermodynamic scaling of the bulk modulus. (a) Bulk modulus $B$ versus $T$ for a range of $P$. The cross, triangle, and square symbols indicate the positions of the onset $T_A$, crossover $T_c$, and glass transition temperatures $T_{\mathrm{g}}$ of glass formation, respectively. (b) $B$ versus $T/\phi^{\gamma}$ with $\gamma =7.0$. Clearly, $B$ does not exhibit thermodynamic scaling.}
\end{figure*}

\begin{figure*}[htb!]
	\centering
	\includegraphics[angle=0,width=0.975\textwidth]{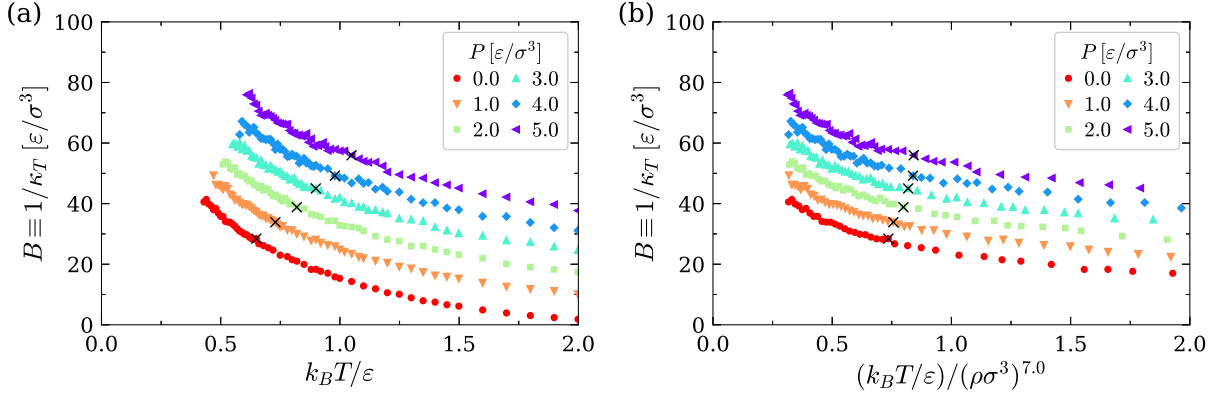}
	\caption{\label{Fig_BT_MD}Simulation test of thermodynamic scaling of the bulk modulus. (a) Bulk modulus $B$ versus $k_BT/\varepsilon$ for a range of $P$. The cross symbols indicate the positions of the onset temperature $T_A$ of glass formation. (b) $B$ versus $(k_BT/\varepsilon) / (\rho \sigma^3)^{\gamma}$ with $\gamma =7.0$. Clearly, $B$ does not exhibit thermodynamic scaling, in accord with the predictions from the GET.}
\end{figure*}

\begin{figure*}[htb!]
	\centering
	\includegraphics[angle=0,width=0.975\textwidth]{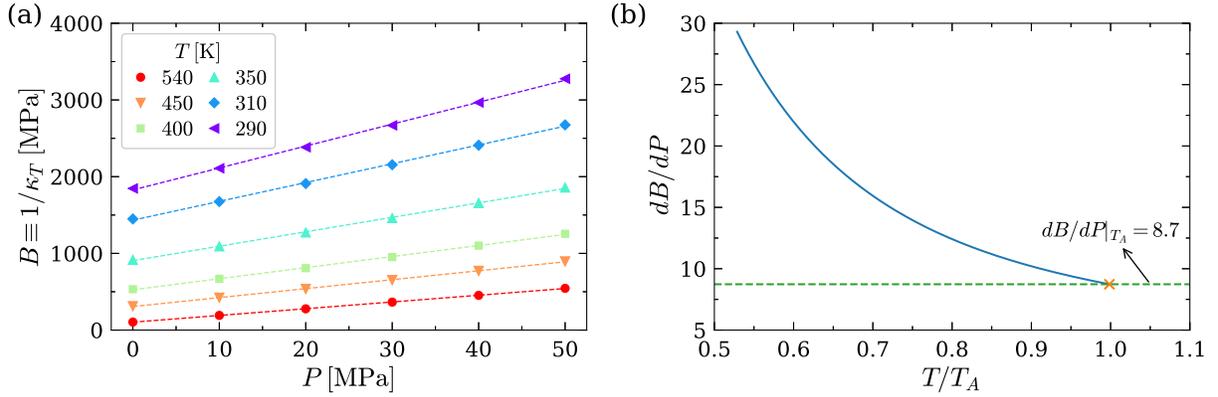}
	\caption{\label{Fig_BP_GET}Pressure derivative of the bulk modulus in the generalized entropy theory. (a) Bulk modulus $B$ versus $P$ for a range of $T$. Lines are linear fits. (b) $dB/dP$ versus $T/T_A$. The cross symbols indicate the position of the onset temperature $T_A$ of glass formation. The dashed line indicates the value of $dB/dP$ at $T_A$, $dB/dP_{T_A} = 8.7$. The derivative of $B$ with respect to $P$ progressively deceases upon heating to a value near the exponent of $7$ for thermodynamic scaling at $T_A$ for Arrhenius relaxation.}
\end{figure*}

\begin{figure*}[htb!]
	\centering
	\includegraphics[angle=0,width=0.975\textwidth]{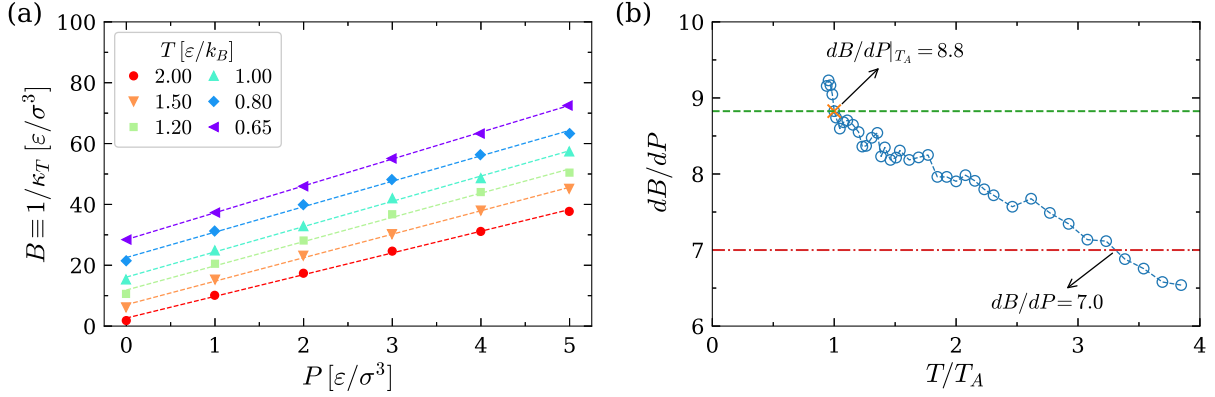}
	\caption{\label{Fig_BP_MD}Pressure derivative of the bulk modulus in simulations. (a) Bulk modulus $B$ versus $P$ for a range of $T$. Lines are linear fits. (b) $dB/dP$ versus $T/T_A$. The cross symbol indicates the position of the onset temperature $T_A$ of glass formation. The dashed line indicates the value of $dB/dP$ at $T_A$, $dB/dP_{T_A} = 8.8$. The derivative of $B$ with respect to $P$ progressively deceases upon heating to a value near the exponent of $7$ in the high $T$ Arrhenius regime. Evidently, $dB/ dP$ is somewhat larger than the scaling exponent at $T_A$, below which non-Arrhenius relaxation starts to be observed. A dash-dotted line indicates that the temperature where the observed thermodynamic scaling exponent is comparable with our $dB/dP$ estimate lies in the Arrhenius regime where $\tau_{\alpha}$ approaches its high frequency limit $\tau_o$.}
\end{figure*}

\begin{figure*}[htb!]
	\centering
	\includegraphics[angle=0,width=0.975\textwidth]{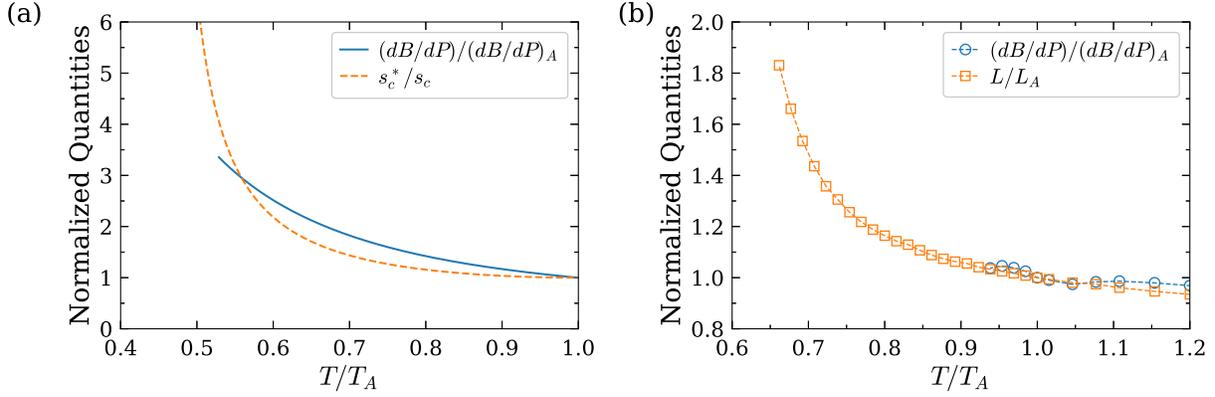}
	\caption{\label{Fig_Compare}Temperature dependence of the pressure derivative of the bulk modulus along with the extent of stringlike cooperative motion. (a) $(dB/dP)/(dB/dP)_A$ and $s_c^*/s_c$ versus $T/T_A$ determined from the generalized entropy theory. (b) $(dB/dP)/(dB/dP)_A$ and $L/L_A$ versus $T/T_A$ determined from simulations.}
\end{figure*}

The GET allows us to consider the molecular origin of configurational anharmonicity quantitatively to better understand thermodynamic scaling since we may calculate the bulk modulus $B$, which is just the reciprocal of the isothermal compressibility, $\kappa_T$. We show our estimates of $B$ versus $T$ in Figure~\ref{Fig_BT_GET} calculated from the GET for the same polymer model for a range of fixed $P$, as described in Section~\ref{Sec_GET}. The $B$ estimates based on the GET and their counterpart based on simulations in Figure~\ref{Fig_BT_MD} clearly do not exhibit thermodynamic scaling, in consistency with the $S(0)$ data shown in Figure~\ref{Fig_S0}, but we may estimate $B$ in the zero pressure limit, $B_o$, and $(dB/dP)_o$, which provides a measure of anharmonicity given that $(dB/dP)_o$ equals $\gamma_M$ in the Murnaghan EOS. In Figures~\ref{Fig_BP_GET} and~\ref{Fig_BP_MD}, we show how $dB/dP$ is estimated in the GET and simulations, along with our predictions for how $(dB/dP)_o$ varies with $T$. As expected, this anharmonicity measure increases progressively upon cooling below $T_A$. This finding accords also with experimental observations of the Gr{\"u}neisen constant of PS as a function of $T$.~\cite{1967_JAP_38_4234} Regardless of the observed universal thermodynamic scaling, the effective anharmonicity $\gamma_M(T) \equiv (dB/dP)_o$ clearly becomes more prevalent in the $T$ regime below $T_A$. For some reason, thermodynamic scaling of relaxation and diffusion remains dependent on $\gamma_M(T_A)$. This finding is a bit of a puzzle, but explains the existence of thermodynamic scaling in the $T$ regime below $T_A$. 

To gain a little further insight into this initially unexpected phenomenon, we plot $dB/dP$ normalized by its value at $T_A$ and $s_c^*/ s_c$ versus $T/T_A$ determined from the GET in Figure~\ref{Fig_Compare}a. The simulation counterpart is provided in Figure~\ref{Fig_Compare}b. Both quantities grow in a parallel fashion, suggesting that the enhanced collective motion in cooled liquids tracks the enhanced anharmonic interactions in the low $T$ regime where standard TST breaks down. It would appear that $\gamma_M(T_A)$ sets the rate of change of $s_c^*/ s_c$ and the corresponding extent of collective motion in cooled liquids. We remark that the fragility of glass formation, which is determined by the differential change of $s_c^*/ s_c$ or the extent of collective motion at a fixed cohesive interaction strength, is then determined by the differential rate of change of the anharmonicity of intermolecular interactions. This provides a novel viewpoint of the origin of fragility in the dynamics of GF liquids and the origin of collective motion. The collective motion is the physical expression of anharmonic interactions and fragility, defined at a fixed cohesive interaction strength. We also note that estimates of $\gamma_M$ over the $T$ range well below $T_A$ can be larger by a factor as large as $3$. This phenomenon probably explains the discrepancy between estimates of $\gamma_M$ from the $P$ derivative of the bulk modulus in the observed thermodynamic scaling exponent in recent measurements by Grzybowski and coworkers.~\cite{2010_PRE_82_013501}

There is a generalization of the Murnaghan EOS that addresses the singular repulsive and attractive contributions of the local interatomic interaction potential, which we have largely ignored in our discussion because we focus on `simple' materials with similar van der Waals interactions governed by an intermolecular potential that has largely the same shape for this broad class of materials. Notably, this is the same starting assumption of Pitzer in his definition of `simple' liquids~\cite{1939_JCP_7_583, 1955_JACS_77_3427, 1955_JACS_77_3433, 2007_JCP_127_224901} and van der Waals in his theory of non-ideal gases,~\cite{Book_Sengers} where the interplay of attractive and repulsive interactions is \textit{essential} in developing an EOS description of non-ideal gases and condensed materials. Of course, the interparticle potential can be rather different in inorganic ionic materials and metals than synthetic polymers, and the potential shape can naturally have its own influence on anharmonicity. Correspondingly, Gilvarry discusses an alternative isothermal EOS based on the assumption that the pair potential can be described by a Mie potential,~\cite{1903_PRE_11_657} in which there is a repulsive power-law potential between interacting particles scaling as a function of distance as $\mathcal{B} / r^n$ and an additive attractive contribution of the opposite sign scaling as $- \mathcal{A} / r^m$, where $\mathcal{A}$ and $\mathcal{B}$ are defined by lattice sums. The original reference to the work by Gr{\"u}neisen is a bit difficult to obtain, but the essential results are reprised by Gilvarry.~\cite{1956_PR_102_331, 1957_JAP_28_1253} This alternative generalization of a lattice theory of solids based on the cubic lattice with this particular family of anharmonic potentials leads to an EOS that generalizes the Murnaghan EOS,
\begin{equation}
	P / [3B_o/(n-m)] = (V_o/V)^{n/3+1} - (V_o/V)^{m/3+1} = (V_o/V)^{n/3+1} [1 - (V_o/V)^{(m-n)/3}],
\end{equation} 
where special potential index pairs have been investigated extensively in applications to particular materials. Interestingly, the Gr{\"u}neisen parameter for this extension of the Debye model can be calculated exactly for this class of model anharmonic potentials~\cite{1956_PR_102_331} and is independent of $\mathcal{A}$ and $\mathcal{B}$, $\gamma_G(\mathrm{Mie}) = (m + n + 3)/6$. More singular attractive and repulsive interactions increase the degree of aharmonicity derived from the shape of the interparticle potential, more singular potential interactions giving rise to larger values of $\gamma_G(\mathrm{Mie})$ and thus $\gamma_M = 2 \gamma_G(\mathrm{Mie}) + 1/3$. For fluids interacting through van der Waals interactions, this source of anhamonicity, i.e., the `potential anharmonicity', probably just makes an additive contribution that is relatively constant in this broad class of materials so that anharmonicity related to monomer shape, molecular stiffness, cohesive interaction strength, etc., `configurational anharmonicity', makes the predominant contribution to the thermodynamics and dynamics of GF liquids arising from anharmonic interactions. Anderson~\cite{1966_PR_144_553} has carefully discussed the assumptions underlying the derivation of this Gr{\"u}neisen-Mie EOS and further derived an often cited analytic expression for the $T$ dependence of the bulk modulus $B$ in terms of anharmonicity parameters based on the same theoretical framework. The resulting equation for $B$ from this analysis is equivalent to the phenomenological `Wachtman equation',~\cite{1966_PR_144_553} which has been highly successful in describing the $T$ dependence of the shear and Young's moduli of diverse inorganic materials.

This type of anharmonicity, which should not be confused with the various forms of anharmonicity that arise in molecular fluids in association with packing and interaction frustration, e.g., due to the presence of semiflexible bonds, could possibly account for thermodynamic scaling if either the attractive or repulsive interactions were formally neglected in comparison to the other term. We anticipate that this approach should have some applicability to certain atomic liquids where the intermolecular potential can reasonably be described by a functional form of this idealized form,~\cite{1971_JCP_55_1128} but we are doubtful that this approach would be very useful in molecular fluids where the interparticle potentials are far more complex. In particular, we expect this framework to provide useful insights into thermodynamic scaling observed in `strongly correlating fluids',~\cite{2008_PRL_100_015701, 2009_JCP_131_234504} where the potential clearly gives rise to potential-virial correlations. Leporini and coworkers~\cite{2016_JCP_145_234904} have considered thermodynamic scaling in liquids described by the class of Mie potentials, providing an attractive set of computational data to analyze from this perspective. We plan to purse an investigation of this kind in the future.

\subsection{\label{Sec_Implication}Implications of Thermodynamic Scaling}

There are many implications of thermodynamic scaling, and we discuss a few here. First, we note that a fixed value of $TV^{\gamma}$ defines a state of fixed configurational entropy density ratio $s_c^*/s_c$ rather than a structurally `isomorphic' state defined by the pair correlation function or its integral. We are not aware of any other thermodynamic property that allows for the definition of such an isokinetic state for the fluid dynamics. The existence of the \textit{fixed reduced isoconfigurational entropy state} means that we may determine $\gamma$ by the simple derivative relation,
\begin{equation}
	\gamma = - (\partial \log T_{\mathrm{g}} / \partial V_{\mathrm{g}})_{P},
\end{equation} 
which has been utilized in some experimental studies to estimate the exponent $\gamma$ for thermodynamic scaling.~\cite{2014_PRL_113_085701} We may deduce relaxation functions from the GET in a reduced variable form, $TV / T_{\mathrm{g}}V_{\mathrm{g}}$, as considered in recent experimental studies.~\cite{2004_EPL_68_58, 2007_JNCS_353_3936, 2011_JCP_135_074901} We plan to discuss this representation based on the GET in a future study.

We may also make significant statements about the $P$ dependence of $T_{\mathrm{g}}$ and $T_m$. The original derivation of a quantitative `Lindemann relation' by Gilvarry~\cite{1956_PR_102_331, 1957_JAP_28_1253} (see also ref~\citenum{1955_PPSB_68_957}) for the melting temperature $T_m$ of crystalline materials was based on the Gilvarry's quantitative, but nonetheless semi-empirical, estimate for crystal instability,~\cite{1956_PR_102_325}
\begin{equation}
	\label{Eq_Tm}
	R T_m = \Omega_{c,m} B_m V_m,
\end{equation} 
where $R$ is the gas constant, $\Omega_{c,m}$ is a `critical constant' that depends on the Poisson ratio, and $B_m$ and $V_m$ are the bulk modulus and molar volume at $T_m$, respectively. Gilvarry~\cite{1956_PR_102_325} attributed the above melting point criterion to an earlier work by Einstein and Lindemann. Equation~\ref{Eq_Tm} offers a significant improvement over Lindemann's original instability condition for melting, based on the assumption that particles actually began to touch, while Gilvarry's criterion, supported by measurements, indicated a critical displacement scale on the order of $1/10$th the interparticle distance.~\cite{1956_PR_102_308, 1956_PR_103_1700} Accordingly, many authors refer to the widely accepted semi-empirical melting criterion given by ref~\ref{Eq_Tm} as the Lindemann-Gilvarry melting criterion.

Although Lindemann and Gilvarry are often recognized for the instability view of melting in the recent scientific literature, we note that it was Sutherland~\cite{1890_PM_30_318, 1891_PM_32_31, 1891_PM_32_215, 1891_PM_32_524} who first formulated an instability of melting as part of his pioneering and ambitious kinetic theory of solids. The theory of Sutherland was the first theory of solids to recognize the significant role of molecular kinetic energy in counterbalancing the strong attractive cohesive interactions characteristic of condensed materials rather than just static `structure' defined by density and molecular configurations in space. The Sutherland theory explained the $T$ dependence of the rigidity of materials as being caused by increasing repulsive intermolecular interactions caused by molecular kinetic energy, as in a heated gas, and his theory of melting, or more generally, fluidization, emphasized that the volume of molecular `domains' explored by the molecules by virtue of their kinetic energy increased in relation to the hard sore size of the molecules to a critical point at which the molecules were no longer `imprisoned' by their neighbors. Sutherland also performed `simulations' with marbles and further suggested that the instability of the solid state might only require a relatively small fraction of the molecules to be mobile in the technical sense. Sutherland recognized that a paradox of this mode of instability is that the change in the density and thus the static dimensions of the metallic materials that he was preoccupied in studying experimentally and theoretically only allowed for an increase of the average interparticle distance by a couple of percent upon changing $T$ from absolute zero to $T_m$ so that the molecular size, as measured by the interparticle distance, remains close to the hard core molecular size. His struggle to understand the relatively small structural change in terms of the density below $T_m$ continues to the present day.~\cite{2010_APL_97_171911} In particular, the density measurements raise the question of how the `domains' explored by the molecules, by virtue of their kinetic energy, could be large enough to explain the existence of a melting instability. Much later with the advent of MD simulation methods, Hoover and coworkers~\cite{1972_JCP_57_1259, 1979_JCP_70_1837} and Reiss and Hammerich~\cite{1986_JPC_90_6252} illustrated the volumes explored by the centers of particles in model liquids, providing insights into the intricate shapes of the `rattle volume' described by Sutherland, and these authors discussed the direct relation of the geometry of this volume to the EOS, as discussed and numerically estimated in more recent works by Sastry et al.~\cite{1998_MP_95_289} and Starr et al.~\cite{2002_PRL_89_125501} for hard-sphere and LJ fluids, respectively.

Given the nearly fixed relation between the molecular size and interparticle distance found in the condensed state, we may recognize Sutherland's model of melting as being basically equivalent to what is termed the Lindermann-Gilvarry criterion of melting. However, the Sutherland criterion appears to have some particular merit in the case of particles, such as polymers that are not spherical, and the Sutherland picture of melting was more recently advocated~\cite{1995_CMS_4_292} without knowledge of Sutherland's pioneering theory of solids and melting. We remark that it is the strong interplay between the thermal energy and interparticle interactions that lies at the core of thermodynamic scaling, and the Sutherland model is the first theory of solids to address the fundamental duality of kinetics and structure in the properties on condensed materials without any reliance on the existence of a presumed lattice structure of the solid. The Sutherland model of solids and its general view of rigidity and mobility as being primarily kinetic rather than structural phenomena have many merits, even from a modern perspective, and this model of the condensed state deserves to be revisited with the aid of simulation studies. Finally, we note the more recent interpretation by Starr and coworkers~\cite{2002_PRL_89_125501} of the loss of ergodicity at the VFT temperature $T_0$ as corresponding to a critical condition in which the cohesive interactions, and corresponding increasing rigidity of cooled liquids, overwhelm the capacity of thermal energy to displace molecules, resulting in particle localization and `structural arrest'. This interpretation of the fundamental nature of solidification accords very well with the Sutherland model of solidification as fundamentally arising from a cessation of capacity of molecules to explore a domain large enough to escape their local environment because of insufficient kinetic energies.

At any rate, the semi-empirical instability criterion by eq~\ref{Eq_Tm} appears to describe $T_m$ for a wide range of materials where $\Omega_{c,m}$ is found to depend on $\gamma_G$. For a fixed Poisson ratio, where the shear modulus $G$ is simply proportional to $B$, we may simply replace $B_m$ in eq~\ref{Eq_Tm} by $G_m$, after modifying $\Omega_{c,m}$ appropriately. The shift of $T_m$ with $P$ is then,~\cite{1956_PR_102_325}
\begin{equation}
	\label{Eq_Tmo}
	T_m / T_{m,o} = (V_{m,o} / V_m)^{\gamma - 1} = (1 + P / P^*)^{(\gamma - 1)/\gamma}.
\end{equation}
Note also that introducing the relation $T_m/T_{m,o} = (V_{m,o}/V_m)^{\gamma - 1}$ into eq~\ref{Eq_Murnaghan} leads to the empirically successful Simon equation where $P$ is the melting pressure, $P_m$.~\cite{1967_PR_161_613, 1956_PR_102_325, 1957_JAP_28_1253, 1963_RMP_35_400} A law of corresponding states analogous to that for gases may be established for solid materials by taking the triple point as the reference temperature in the Simon equation.~\cite{1956_PR_102_325}

Arguments have also been given that the glass transition should be governed by a Lindemann relation,~\cite{2008_ACP_137_125, 2010_APL_97_171911} and by extension of eq~\ref{Eq_Tm}, we adopt the corresponding relation for $T_{\mathrm{g}}$,
\begin{equation}
	R T_{\mathrm{g}} = \Omega_{c,\mathrm{g}} B_{\mathrm{g}} V_{\mathrm{g}}.
\end{equation}
Egami and coworkers derived an equation of the general form of eq~\ref{Eq_Tm} for $T_{\mathrm{g}}$ with the specification of the Poisson ratio dependent prefactor for metallic GF materials and observed good consistency with experiment.~\cite{2007_PRB_76_024203} We also notice proportionalities between $T_{\mathrm{g}}$ and $T_m$ and the bulk and shear moduli and with each other with a constant of proportionality depending on the Poisson ratio. This has been verified in experimental studies.~\cite{2012_JCP_136_224108, 2015_JNCS_407_14}

By the same argument as above for $T_m$, we may expect $T_{\mathrm{g}}$ to exhibit the same type of 
$P$ dependence as for $T_m$,
\begin{equation}
	\label{Eq_Tgo}
	T_{\mathrm{g}} / T_{\mathrm{g}, o} = (V_{m,o} / V_m)^{\gamma - 1} = (1 + P / P^*)^{(\gamma - 1)/\gamma}.
\end{equation}
Following Andersson and Andersson,~\cite{1998_Mac_31_2999} we have fitted the $P$ dependence of all the characteristic temperatures ($T_A$, $T_c$, $T_{\mathrm{g}}$, and $T_0$) of glass formation to this general functional form in our past works,~\cite{2016_MacroLett_5_1375, 2017_Mac_50_2585} along with $\Delta H_A$. We can now see the theoretical rationale for this type of relationship. 

Finally, we may gain physical insight into the $P$ dependence of the thermal expansion coefficient $\alpha_P$. We can evidently calculate $\alpha_P$ from the GET, but the physical meaning of the result is not always clear. For temperatures above the Debye temperature, a reasonable assumption for many fluids is that the vibrational degrees of freedom are activated. We may invoke the approximation of Delong and Petit,~\cite{1819_ACP_10_395, Book_Mcquarrie} $C_V \approx 3 R$, so that eq~\ref{Eq_Tm} implies,~\cite{2009_PM_89_1757}
\begin{equation}
	\label{Eq_DP}
	\alpha_P \approx \gamma_G (3 R) \kappa_T / V.
\end{equation}
This indicates that we may estimate the change of $\alpha_P$ with $P$ as, 
\begin{equation}
	\label{Eq_alphao}
	\alpha_P / \alpha_{P,o} \approx (1 + P / P^*)^{- \delta_T},
\end{equation} 
where the exponent $\delta_T$ has been termed the Anderson-Gr{\"u}nesisen parameter.~\cite{1967_JGR_72_3661, 1996_PRB_54_7034, 1993_JCP_99_5369} This relation, with $\delta_T$ being an adjustable parameter, has been used in applications in geophysical science.~\cite{Book_Anderson} Given eqs~\ref{Eq_Tmo},~\ref{Eq_Tgo}, and~\ref{Eq_alphao}, we can then readily understand the near constancy of $\alpha_P (T_m) T_m$ and $\alpha_P(T_{\mathrm{g}}) T_{\mathrm{g}}$ in both crystalline and GF materials.~\cite{2010_APL_97_171911, 1962_JCP_37_1003} This is just the result of a near cancellation of anharmonic interaction effects in $\alpha_P$ and the characteristic temperatures, $T_m$ and $T_\mathrm{g}$. As a remarkable general property of condensed materials, we note that there is a general approximate inverse relation between the material stiffness (tensile modulus) of diverse condensed materials and the square of $\alpha_P$,~\cite{1967_JAP_38_4234} although there are exceptions of this general trend in some polymeric materials. We also mention that Ledbetter~\cite{1994_PSS_181_81} has derived a very interesting interrelation between the thermal expansion coefficient and the bulk modulus and it derivative with respect to $T$. 

The `anomalous' character of thermal expansion in polymer materials can be rationalized, at least in part, by the observation that approximating $C_V$ by a constant is often not suitable in polymer materials because the potential energy has large contributions arising from the intramolecular bonds, in addition to the intermolecular interactions similar to those dominating the potential energy in atomic and small-molecule liquids.~\cite{1963_JAP_34_107, 1967_JAP_38_4234} Correspondingly, the most effective models of the specific heat of polymer materials have been modeled by taking the polymer chains to be one-dimensional crystals whose interchain interactions are modeled by the Debye model or its extension to include anharmonicity.~\cite{1963_JAP_34_107, 1967_JAP_38_4234} The application of this type of model to many experimental systems has indicated that the Debye temperature associated with chain bonding tends to be quite large so that the $T$ regime where $C_V$ can be taken as just a constant as in eq~\ref{Eq_DP} is limited in these materials. We have recently discussed these issues regarding the complexity of the $T$ dependence of $C_V$ in polymers from a computational standpoint.~\cite{2016_Mac_49_8341}

Thermodynamic scaling thus provides quantitative and qualitative insights into widely observed scaling relations for how the melting temperature and the characteristic temperatures of glass formation depend on pressure.

\section{Summary}

We have shown that thermodynamic scaling can be derived within a transition state theory framework in conjunction with the well-validated Murnaghan equation of state~\cite{1944_PNAS_30_244, Book_Murnaghan, 1995_IJT_16_1009} relating changes of pressure to changes in volume in condensed materials and the Gilvarry interpretation of this equation of state~\cite{1956_PR_102_331, 1957_JAP_28_1253} in terms of an anharmonic extension of the Debye model of solid materials that incorporates Gr{\"u}neisen's quantification of material anharmonicity in interatomic interactions.~\cite{1955_JCP_23_1925} The strict applicability of these calculations is limited to the Arrhenius regime where the fluid may be considered dynamically homogeneous so that relaxation and diffusion can be appropriately described by transition state theory. We have addressed the important extension to the temperature regime below the onset temperature $T_A$ for non-Arrhenius relaxation and diffusion through the vehicles of the generalized entropy theory and molecular dynamics simulation, where the same thermodynamic scaling has been shown to hold below $T_A$ as in the Arrhenius regime. This fact has been rationalized by the observation of thermodynamic scaling of the configurational entropy density $s_c$, normalized by its high temperature value $s_c^*$, in explicit calculations based on the generalized entropy theory. We have also shown that the extent $L$ of stringlike collective motion normalized by its value $L_A$ at $T_A$, which determines the temperature dependence of the activation energy below $T_A$ in the string model, obeys thermodynamic scaling, consistent with the identification of the strings with the cooperatively rearranging regions of Adam and Gibbs.~\cite{1965_JCP_43_139}

As noted by Dyre and coworkers.~\cite{2009_JCP_131_234504} the existence of thermodynamic scaling offers a `filter' for assessing models of glass formation. Accordingly, we found that the static structure factor in the long wavelength limit, $S(0)$, and the isothermal compressibility, do not exhibit thermodynamic scaling, raising questions about theories that purport that the structural relaxation time and diffusion are directly related to $S(0)$.~\cite{Book_Gotze, 2004_JCP_121_1984, 2004_JCP_121_2001, 2010_ARCMP_1_277, 2014_JCP_140_194506, 2014_JCP_140_194507, 2015_Mac_48_1901, 2016_Mac_49_9655} On the other hand, we found that the mean square displacement at a caging time, the Debye-Waller parameter $\langle u^2 \rangle$, follows thermodynamic scaling to a high approximation, a point made by Leporini and coworkers~\cite{2008_NaturePhys_4_42} and others.~\cite{2011_JCP_135_164510} This means that the localization model of glass formation,~\cite{2012_SoftMatter_8_11455, 2015_PNAS_112_2966, 2016_JSM_054048} and the somewhat related model of Leporini and coworkers,~\cite{2008_NaturePhys_4_42} passes this basic consistency test associated with thermodynamic scaling. Thermodynamic scaling then offers a powerful tool for testing models of glass formation, although this scaling property in itself does not establish any such theory.

Thermodynamic scaling also has great potential for providing new research directions on glass-forming liquids. The finding that $\langle u^2 \rangle$ and the reduced configurational entropy obey a common thermodynamic scaling has naturally led us to consider whether these properties might have some direct relationship. Based on the further plausible assumption that $\langle u^2 \rangle^{3/2}$ might be interpreted as local `compressibility', we have found based on the GET that there is indeed a direct relation between the fluid isothermal compressibility and configurational entropy once both quantities are properly reduced. This scaling relation provides insight into why the isothermal compressibility and $S(0)$ do not obey thermodynamic scaling, and more importantly, it allows the entropy theory to be recast in terms of a reduced isothermal compressibility, a quantity that is directly measurable. This result provides conceptual insight into the interrelation between changes in the material rigidity and configurational entropy that lie at the heart of `shoving' or elastic models,~\cite{2006_RMP_78_953, 2012_JCP_136_224108, 2015_JNCS_407_14} and the localization model,~\cite{2012_SoftMatter_8_11455, 2015_PNAS_112_2966, 2016_JSM_054048} a type of `dynamic free volume' model of glass formation that emphasizes the volume explored by particles in their condensed state rather than the local density as in Sutherland's kinetic theory of solids.~\cite{1890_PM_30_318, 1891_PM_32_31, 1891_PM_32_215, 1891_PM_32_524}

We have considered the implications of our newly found capacity to quantify the aharmonicity in condensed materials and found that there are different types of anharmonicity, namely, potential and configurational anharmonicities, derived from the shape of the pair potential and from the monomer shape, rigidity, connectivity, and other factors that influence molecular packing, respectively. This allows us to understand why the thermodynamic scaling exponent varies greatly in polymeric and other molecular materials in which molecules interact through a similar local pair potential describing their van der Waals interactions. The generalized entropy theory allows for a direct calculation of how the thermodynamic scaling exponent $\gamma$ relates to the fragility of polymeric glass-forming materials,~\cite{2013_JCP_138_234501} along with other thermodynamic and dynamic properties of these materials within a consistent theoretical framework. We have also found the unexpected result that the Murnaghan exponent $\gamma_M$, our anharmonicity measure, can change appreciably with temperature and that these changes track the extent of stringlike collective motion below the onset temperature $T_A$ for non-Arrhenius relaxation. This phenomenon, which certainly requires further investigation, seems to suggest that the collective motion below $T_A$ is a direct physical expression of a growing anharmonicity of intermolecular interactions in cooled liquids. Anharmonicity is either the cause or the effect of this form of dynamic heterogeneity.

Finally, after establishing thermodynamic scaling as an attribute of our simulation observations on glass-forming polymer melts and the generalized entropy theory and the string model of glass formation, we have considered some of the many ramifications of thermodynamic scaling, such as the shift of the glass transition and melting temperatures with pressure and other material properties (e.g., thermal expansion coefficient and isothermal compressibility) that depend strongly on anharmonic interactions. We have shown that many phenomenological relationships, such as the Simon relation~\cite{1956_PR_102_325} for the pressure dependence of the melting temperature and its analog for the glass transition temperature of glass-forming materials, the Anderson relation~\cite{1967_JGR_72_3661, 1996_PRB_54_7034, 1993_JCP_99_5369} for the thermal expansion coefficient, the Tait equation,~\cite{1976_JAP_47_5201} and the Bridgeman equation,~\cite{1955_JCP_23_1925, 1966_RMP_38_669, 1966_JPCS_27_547} can be understood within the unifying framework of the Murnaghan equation of state. We have also obtained a better understanding of the near constancy of the product of the thermal expansion coefficient and the melting or glass transition temperatures in materials that solidify through crystallization or glass formation.

\begin{acknowledgement}
W.-S.X. acknowledges the support from the National Natural Science Foundation of China (No. 21973089). A portion of the simulations was conducted at the Oak Ridge National Laboratory's Center for Nanophase Materials Sciences, which is a DOE Office of Science User Facility. This research used resources of the Compute and Data Environment for Science (CADES) at the Oak Ridge National Laboratory, which is supported by the Office of Science of the U.S. Department of Energy under Contract No. DE-AC05-00OR22725. This research also used resources of the Network and Computing Center at Changchun Institute of Applied Chemistry, Chinese Academy of Sciences.
\end{acknowledgement}


\bibliography{refs}

\end{document}